\documentstyle[12pt,preprint,aps,graphics]{revtex}
\def\bea{\begin{eqnarray}}
\def\eea{\end{eqnarray}}
\def\be{\begin{equation}}
\def\ee{\end{equation}}
\begin{document}
\tighten
\title{Compositional and orientational ordering in
  rod-coil diblock copolymer melts} \author{Manuel
  Reenders\footnote{m.reenders@chem.rug.nl} and Gerrit ten Brinke}
\address{
  Department of Polymer Chemistry and Materials Science Center,\\
  University of Groningen, Nijenborgh 4, 9747 AG Groningen, The
  Netherlands} 
\date{\today} 
\maketitle
\begin{abstract}
  The phase behavior of a melt of monodisperse rod-coil diblocks is
  studied.  We derive a Landau free energy functional for both a
  compositional and a nematic order parameter.  The excluded volume
  interaction between the rod blocks is modeled by an attractive
  Maier-Saupe interaction.  The incompatibility between rod- and coil
  blocks is modeled by the usual Flory-Huggins interaction.  For a
  large volume fraction of the rods, a transition from isotropic to
  nematic to smectic C is observed upon decreasing the temperature,
  whereas for small rod volume fraction, spherical, hexagonal, and
  lamellar structures prevail. In the smectic C phase, the rod
  orientation angle with respect to the lamellar normal increases
  rapidly from $35$ to $40^o$ close to the nematic/smectic-C phase
  boundary to values between $45$ and $55^o$.
\end{abstract}
%
%
\section{Introduction}
Due to chemical incompatibility flexible AB diblocks are known to
phase separate on a microscopic scale characterized by the radius of
gyration of the blocks.  The morphology of the microphases in the
melt is successfully described in various approaches, which treat the AB
diblocks as Gaussian chains and the incompatibility by a 
Flory-Huggins interaction \cite{le80,se85,masc94,bafr90}.
Rod-coil diblock copolymers are a special class of AB diblock
copolymers.  Besides the usual incompatibility between distinct blocks, 
the liquid crystalline behavior of the stiff blocks can 
have a profound effect on the phase behavior.
The combination of phase separation and orientational ordering gives rise to 
a wealth of different phases.
In this paper a theoretical description of both phase separation 
and orientational ordering is developed.

In experiments on various rod-coil or liquid crystalline polymeric
systems interesting phase behavior and micro-structures are observed 
\cite{rast94,racast97,lechohzi01,lekihwkiohzi01,sczayafith00,chthobma96,muobth97,warogo01,elclpete00,bert}.
Linear rod-coil multi-block oligomers can have orientationally ordered lamellar, 
hexagonal, spherical (bcc), and bicontinuous cubic 
liquid crystalline structures when the 
rod volume fraction is less than the coil 
fraction \cite{rast94,racast97,lechohzi01,lekihwkiohzi01}.
Schneider {\em et al.} \cite{sczayafith00}
reported  results on a particular rod-coil diblock copolymer
melt in which microphase separation occurs for higher temperatures
than the liquid-crystalline order; although at high temperatures the microphase is
bcc or gyroid, whenever anisotropic orientational order starts to play a role, 
the lamellar or smectic phase appears.
One of the most striking features is the appearance of smectic C phases, 
and so-called zig-zag and arrowhead phases \cite{chthobma96,muobth97}
for intermediate rod volume fractions.

The theoretical description of the phase behavior of rod-coil diblock
copolymers requires the introduction of a compositional (density)
order parameter and a (anisotropic) orientational order parameter or
nematic order parameter.  In the work of Holyst and Schick
\cite{hosc92b}, the interactions and the excluded volume effects such
as the steric repulsion between the rod blocks are treated in an
approximate manner via the Flory-Huggins interaction, Maier-Saupe
interaction, and the incompressibility constraint.  The stiff part of
the diblocks can be modeled by rigid rods \cite{hosc92b}, wormlike
chains \cite{lifr93} or freely jointed rods \cite{sigolifr94}, having
a Maier-Saupe interaction \cite{hoos01}.  The isotropic-nematic
transition in various polymer systems was studied in a number of
papers \cite{lifr93,frle90,hosc92a,hagozvsefr01}.  Using
self-consistent field theory, the interplay between phase separation
and liquid crystalline order has been studied ignoring morphologies
other than lamellar \cite{nesc96,maba98}.
Gurovich found that copolymer melts can order in four distinct ways under the influence of external
ordering fields ({\em e.g.}, electric fields) close to the spinodal
point \cite{gu95}.

For large rod volume fractions, the smectic behavior of layered
rod-coil diblock copolymers in the strong segregation limit was
studied \cite{seva86,ha89,sesu92,maba98}. For small rod volume
fractions, so-called hockey puck phases might appear in the strong
segregation region \cite{wifr92} or whenever phase separation occurs
prior to nematic ordering \cite{wiha93}. Recently, a two-dimensional
dynamic mean field model using the Maier-Saupe interaction was
employed for a melt of semiflexible polymers \cite{hagozvsefr01}.
Also the time evolution of morphology formation was studied in two
dimensions for liquid-crystal/polymer mixtures showing a large
influence of nematic ordering on phase separation \cite{lagllali99}.
Moreover, in two dimensions, the phase behavior of rod-coil diblock
copolymer melts was obtained using a self-consistent field lattice
model \cite{lige01}. Interestingly, evidence for the existence of
metastable zigzag structures which have been observed in experiments
\cite{chthobma96,muobth97}, was found \cite{lige01}.

In this paper, along the lines of Holyst and Schick \cite{hosc92b} and
Singh {\em et al.} \cite{sigolifr94}, a monodisperse incompressible
rod-coil diblock copolymer melt is studied with both a Flory-Huggins
interaction and a Maier-Saupe interaction.  The Maier-Saupe
interaction models the steric repulsion between the rods, and
henceforth favors alignment.  Special attention is paid for the
interplay between nematic ordering and microphase separation. One of
the main questions addressed is: ``How does the nematic ordering
affect the microphase separation?''  The Landau free energy is
calculated up to the fourth order in two order parameters in the weak
segregation limit.  A density order parameter describes the tendency
to microphase separation, whereas a nematic order parameter describes
anisotropic alignment of rods in the melt. For blends, the free energy
up to fourth order was already obtained by Liu and Fredrickson
\cite{lifr93}. Perhaps another way of addressing the current issue is
using the density functional formalism, {\em e.g.}, see the paper by
Fukuda and Yokoyama \cite{fuyo01} and references therein.

The setup of the present paper is as follows.
In the next section a description of the partition function
describing the rod-coil diblock melt is given, followed
by an introduction into 
the Landau mean field expansion. 
Subsequently, the free energy expansion
up to second order is briefly reviewed, illustrating the instability
of the isotropic phase with respect to nematic ordering 
and microphase separation.
Then, the actual Landau free energy up to the
fourth order is derived in the so called first harmonics approximation (FHA).
The minimization of the free energy and the corresponding phase diagrams are
discussed.
In Appendix~\ref{ap_gamma2} the second order Landau coefficient 
or inverse ``scattering matrix'' is computed. 
Finally in Appendix~\ref{ap_vert} and \ref{ap_coilrodfies} 
the pertinent three- and four-point single 
chain correlation functions and vertices are given.

\section{The model}\label{sec_mod}
We consider an incompressible melt of $n$ 
monodisperse rod-coil diblocks in volume $V$.
The flexible (coil) part of the diblock is modeled by a Gaussian chain,
and the rigid part by a ``thin'' rod.
Each diblock molecule consists of $N$ segments 
of which $N_C=f_C N$ are coil segments and $N_R=f_RN$ are rod segments,
with $f_R=(1-f_C)$.
Thus, the volume fraction of the coil part is given by $f_C=N_C/N$.
We keep the total density constant, $\rho_0=nN/V$.
The characteristic length scale of the coil block 
is given by the radius of gyration, 
$R_g=b\sqrt{N_C/6}$, with $b$ the statistical coil segment length scale.
The length of the rod is characterized by $\ell=N_R b^\prime$, with
$b^\prime$ the rod segment length.
In this paper, we set $b=b^\prime=1$, thus $\rho_0=nN/V=1/b^3=1$.

The Hamiltonian for our melt contains two terms:
one describing the standard Flory-Huggins repulsion between 
the rod and coil blocks and a second term describing orientational ordering
of the rod blocks in the melt.
The Hamiltonian for our model is
\bea
H_I=\chi\int d\vec x\,\rho_R(\vec x)\rho_C(\vec x)
-\frac{\omega}{2}\int d\vec x\,
S^{\mu\nu}_R(\vec x)S^{\mu\nu}_R(\vec x),
\qquad (\mu,\nu=1,\dots,3)
\label{hamil}
\eea 
where $\rho_C$ and $\rho_R$ are the densities of the coil and the
rod blocks, respectively, and $S^{\mu\nu}_R$ is the nematic order
parameter tensor.  The tensor $S^{\mu\nu}_R$ is symmetric and
traceless.  $\chi$ is the usual Flory-Huggins parameter.  The second
term in Eq.~(\ref{hamil}) is the Maier-Saupe interaction with the
parameter $\omega$.  It effectively describes the excluded volume
interaction between the rod blocks in the melt favoring their
alignment.

\begin{figure}
\resizebox*{0.6\columnwidth}{!}{\includegraphics{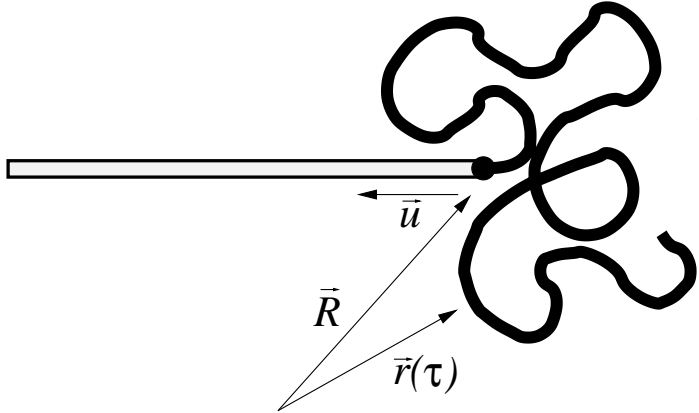}}
\caption{Parametrization of the configuration of a rod-coil diblock.}
\label{fig_rc1}
\end{figure}

The conformation of a diblock in the melt is given by the vector 
function $\vec r(\tau)$ describing the contour of the coil, the vector $\vec R$
giving the position of the joint, and the unit vector $\vec u$ describing 
the orientation of the rod, see Fig.~\ref{fig_rc1}.
Then, the (single-chain) partition function 
of a rod-coil diblock in external fields is 
\bea
&&Q[w_R, w_C,W^{\mu\nu}]=C\int {\cal D}\vec r d\vec R d\vec u\,
\delta(\vec r(f_C)-\vec R)\delta(|\vec u|-1)\nonumber\\
&&\times \exp{\left\{-\frac{3}{2Nb^2}\int_0^{f_C} d\tau\,
|{\dot {\vec r}}|^2 +\int w_R\hat\rho_R
+\int w_C\hat\rho_C+\int W^{\mu\nu}\hat S_R^{\mu\nu}\right\}}.\label{Qscdef}
\eea
where $W^{\mu\nu}$ is an external tensor field which couples 
to the nematic-order operator.
The normalization constant $C$ is chosen such that $Q[0,0,0]=1$.
The density and orientational operators are
\bea
\hat\rho_{R}(\vec x)&=&\frac{f_RN}{\ell}\int\limits_0^{\ell}ds\,
\delta(\vec x-\vec R -s\vec u),\label{densrdef}\\
\hat \rho_{C}(\vec x)&=&N\int\limits_0^{f_C}ds\,
\delta(\vec x-\vec r(s)),\label{denscdef}\\
\hat S_{R}^{\mu\nu}(\vec x)&=&\frac{f_RN}{\ell}\int\limits_0^{\ell}ds\,
\delta(\vec x-\vec R-s\vec u)\left[u^\mu u^\nu
-\frac{1}{3}\delta^{\mu\nu} u^2\right].\label{snemdef}
\eea 
From Eqs.~(\ref{Qscdef})-(\ref{snemdef}) all single chain correlation
functions can be obtained by differentiating 
$Q$ with respect to the external fields.

The partition function of the whole incompressible melt can be described
by
\bea
{\cal Z}&=&\left\{\prod_{m=1}^n\int {\cal D}\vec r_m
d\vec R_m d\vec u_m\,{\cal P}[\{\vec r_m, \vec R_m,\vec u_m\}]
\right\}
\delta(1-\hat\rho_R(\vec x)-\hat\rho_C(\vec x))\nonumber\\
&\times&
\exp{\left\{-\chi\int d\vec x\,\hat \rho_R(\vec x)\hat\rho_C(\vec x)
+\frac{\omega}{2}\int d\vec x\,\hat S^{\mu\nu}_R(\vec x)
\hat S^{\mu\nu}_R(\vec x)\right\}},
\eea
where 
\bea
{\cal P}[\{\vec r_m, \vec R_m,\vec u_m\}]=
C\delta(\vec r_m(f_C)-\vec R_m)\delta(|\vec u_m|-1)
\exp{\left\{-\frac{3}{2Nb^2}\int_0^{f_C} d\tau\,
|{\dot {\vec r_m}}|^2 \right\}}.\label{Pdef}
\eea
After a number of Legendre transformations,
we obtain 
\bea
{\cal Z}&\propto&
\int {\cal D}\psi_R{\cal D}\psi_C{\cal D}S^{\mu\nu}_R
{\cal D}J_R{\cal D}J_C{\cal D}J^{\mu\nu}\,
\delta(\psi_R+\psi_C)\nonumber\\
&\times&\delta\left(\int \psi_R\right)
\delta\left(\int \psi_C\right)
\delta\left(\int J_R\right)
\delta\left(\int J_C\right)
\nonumber\\
&\times&
\exp{\left\{-\chi\int\psi_R\psi_C
+\frac{\omega}{2}\int
S^{\mu\nu}_R S^{\mu\nu}_R\right\}}
\nonumber\\
&\times&
\exp{\left\{
-\int J_R\psi_R
-\int J_C\psi_C-\int J^{\mu\nu}S^{\mu\nu}_R-G[J_R,J_C,J^{\mu\nu}]
\right\}},
\label{legpartZ}
\eea
where
\bea
G[J_R,J_C,J^{\mu\nu}]&=&-n\ln Q[J_R,J_C,J^{\mu\nu}],
\eea
and where we have introduced the fields
\bea
\psi_C(\vec x)=\rho_C(\vec x)-f_C,\qquad 
\psi_R(\vec x)=\rho_R(\vec x)-f_R. 
\eea
Making use of the above definition of $G$,
the Landau mean field free energy can be obtained as an expansion
in powers of the concentration profiles $\psi_R$, $\psi_C$ and the nematic
order parameter $S_R^{\mu\nu}$.
\section{The Landau mean-field free energy}\label{sec_lamf}
We can write the ``entropic'' and ``interaction'' parts, respectively,
of Eq.~(\ref{legpartZ}) as follows:
\bea
\exp{\left\{-F_I[\phi]\right\}}&\equiv&\exp{\left\{
-I_{\alpha\beta}\int\phi_\alpha\phi_\beta\right\}},
\\
\exp{\left\{-F_S[\phi]\right\}}&\equiv&
\int {\cal D}J\,\exp{\left\{
-\int J_\alpha\phi_\alpha-G[J]\right\}},
\eea
with the shorthand notation, 
\bea
\phi_\alpha&=&(\psi_R,\psi_C,S^{\mu\nu}_R),\\
J_\alpha&=&(J_R,J_C,J^{\mu\nu}),\\
\hat\rho_\alpha&=&(\hat\rho_R,\hat\rho_C,\hat S^{\mu\nu}_R).
\eea
Thus the Greek index $\alpha=R,C,S$.
So that
\bea
{\cal Z}=\exp{\left\{-{\cal F}\right\}}\propto \int {\cal D}\phi\,\exp{\left\{-F[\phi]\right\}},\qquad
F=F_I+F_S.
\eea
The entropic part of the Landau free energy reads
\bea
\frac{NF_S}{V}&=&
\frac{1}{2!V^2}\sum_{k_1}
\phi_\alpha(k_1)\phi_\beta(-k_1)\Gamma^{(2)}_{\alpha\beta}(k_1)\nonumber\\
&+&\frac{1}{3!V^3}\sum_{k_1}\sum_{k_2}
\phi_\alpha(k_1)\phi_\beta(k_2)\phi_\gamma(-k_1-k_2)
\Gamma^{(3)}_{\alpha\beta\gamma}(k_1,k_2)
\nonumber\\
&+&\frac{1}{4!V^4}\sum_{k_1}\sum_{k_2} \sum_{k_3} 
\phi_\alpha(k_1)\phi_\beta(k_2)\phi_\gamma(k_3)\phi_\delta(-k_1-k_2-k_3)\nonumber\\
&\times&\Gamma^{(4)}_{\alpha\beta\gamma\delta}(k_1,k_2,k_3).
\label{lanentr}
\eea
The vertices $\Gamma$ are
\bea
\Gamma^{(2)}_{\alpha\beta}(k_1)&=&
\left[W^{(2)}_{\alpha\beta}(k_1)\right]^{-1},\label{gam2abdef}\\
\Gamma^{(3)}_{\alpha\beta\gamma}(k_1,k_2)&=&
-\Gamma^{(2)}_{\alpha\alpha^\prime}(k_1)
\Gamma^{(2)}_{\beta\beta^\prime}(k_2)
\Gamma^{(2)}_{\gamma\gamma^\prime}(-k_1-k_2)
W^{(3)}_{\alpha^\prime\beta^\prime\gamma^\prime}(k_1,k_2),
\eea
and
\bea
\Gamma^{(4)}_{\alpha\beta\gamma\delta}(k_1,k_2,k_3)&=&
-\Gamma^{(2)}_{\alpha\alpha^\prime}(k_1)
\Gamma^{(2)}_{\beta\beta^\prime}(k_2)
\Gamma^{(2)}_{\gamma\gamma^\prime}(k_3)
\Gamma^{(2)}_{\delta\delta^\prime}(-k_1-k_2-k_3)\nonumber\\
&\times& \bigg[W^{(4)}_{\alpha^\prime
\beta^\prime\gamma^\prime\delta^\prime}(k_1,k_2,k_3)
-\delta_K(k_1+k_2)W^{(2)}_{\alpha\beta}(k_1)W^{(2)}_{\gamma\delta}(k_3)\nonumber\\
&-&\delta_K(k_1+k_3)W^{(2)}_{\alpha\gamma}(k_1)
W^{(2)}_{\beta\delta}(k_2)
-\delta_K(k_2+k_3)W^{(2)}_{\alpha\delta}(k_1)W^{(2)}_{\beta\gamma}(k_2)
\nonumber\\
&-&W^{(3)}_{\alpha^\prime\beta^\prime \mu}(k_1,k_2)\Gamma^{(2)}_{\mu\nu}(-k_1-k_2)
W^{(3)}_{\nu\gamma^\prime\delta^\prime}(k_1+k_2,k_3)\nonumber\\
&-&W^{(3)}_{\alpha^\prime\gamma^\prime \mu}(k_1,k_3)\Gamma^{(2)}_{\mu\nu}(-k_1-k_3)
W^{(3)}_{\nu\beta^\prime\delta^\prime}(k_1+k_3,k_2)\nonumber\\
&-&W^{(3)}_{\alpha^\prime\delta^\prime \mu}(k_1,-k_1-k_2-k_3)
\Gamma^{(2)}_{\mu\nu}(k_2+k_3)
W^{(3)}_{\nu\gamma^\prime\beta^\prime}(-k_2-k_3,k_3)
\bigg].\label{gam4ppppdef}
\eea
The functions $W^{(n)}$ are 
the single chain correlation functions
\bea
W^{(n)}_{\alpha_1 \alpha_2\cdots\alpha_n}
(k_1,k_2,\cdots,k_{n-1})\equiv N^{-n}\langle \hat \rho_{\alpha_1}(k_1)\hat
\rho_{\alpha_2}(k_2)\cdots\hat\rho_{\alpha_n}(-\sum_{i=1}^{n-1}k_i)\rangle_0,
\eea
where the average $\langle\cdots\rangle_0$ is defined as
\bea
\langle {\cal F}[\{\vec r, \vec R,\vec u\}]\rangle_0\equiv 
\int {\cal D}\vec r d\vec R d\vec u\,{\cal P}[\{\vec r, \vec R,\vec u\}]
{\cal F}[\{\vec r, \vec R,\vec u\}],
\eea
where ${\cal P}$ is given by Eq.~(\ref{Pdef}).
\section{The spinodal}\label{sec_spin}
First, we consider the Landau expansion up to second order
in the density and nematic order parameters,
to analyze the instability of the isotropic phase
with respect to microphase separation and nematic ordering.
Therefore, following the approach of
Holyst and Schick \cite{hosc92b} 
and Singh {\em et al.} \cite{sigolifr94},
we consider the free energy
\bea
{\cal F}[\psi,S]=\frac{1}{2}\sum_{\vec q}(\psi(-\vec q),S(-\vec q)) 
\Gamma^{(2)}(\vec q) 
\left(\begin{array}{c}\psi(\vec q)\\S(\vec q)\end{array}\right),
\eea
where $\Gamma^{(2)}$ is a $2\times 2$ ``scattering matrix''.
This matrix $\Gamma^{(2)}$ can be obtained from
the matrix ${\bf \Gamma}$ given in Eq.~(\ref{Gmatdef}) 
in Appendix~\ref{ap_gamma2};
\bea
\Gamma^{(2)}=\left(\begin{array}{cc}
h_{RR}+h_{CC}-2h_{RC}-2N\chi& 2(h_{RS}-h_{CS})/3\\
2(h_{RS}-h_{CS})/3& 
\bar h_{SS}-2N\omega/3
\end{array}\right),
\eea
where 
\bea
\bar h_{SS}=\sum_{i=1}^3 
h_{SSi}\left(\frac{q^\mu q^\nu}{q^2}-\frac{\delta^{\mu\nu}}{3}\right)
\left(\frac{q^\rho q^\sigma}{q^2}-\frac{\delta^{\rho\sigma}}{3}\right)
T^{\mu\nu\rho\sigma}_i(q),
\eea
see Appendix~\ref{ap_gamma2}.
For the nematic order tensor $S_R^{\mu\nu}$ we have taken the Ansatz
\bea
S^{\mu\nu}_R(\vec q)=S(q)
\left(\frac{q^\mu q^\nu}{q^2}-\frac{\delta^{\mu\nu}}{3}\right),
\eea
where $S$ is a scalar function.
The nematic ordering is taken parallel to the wave vector $\vec q$.

The ``scattering matrix'' $\Gamma^{(2)}$ depends on $\vec q$, $f_R$,
$N$ and the ratio $r=\omega/\chi$.  The spinodal is determined by the
root of the determinant of $\Gamma^{(2)}$; \bea \det
\Gamma^{(2)}=0\quad \Longrightarrow \quad \chi(\vec q,f_R,N,r).  \eea
The real spinodal is given by the global minimum of $\chi$ 
({\em i.e.}, the highest $T$) with respect to $\vec q$.  Whenever one of the
eigenvalues (or both of them) of $\Gamma^{(2)}$ becomes negative, the
isotropic phase becomes unstable.  The onset of the nematically
ordered phase is given by the minimum of the root $\chi$
at wave vector $q=0$.  The onset of microphase separation is given by
the minimum of the root $\chi$ at some nonzero $q=q^\ast$, which gives
the inverse characteristic length scale of the microphase.  The
minimum at $q=0$ for nematic ordering gives rise to a global nematic
order parameter $S(\vec q)\rightarrow S$.  All this is discussed in
detail by Singh {\em et al.} \cite{sigolifr94}, although for slightly different
microscopic models.  The situation where spinodal instabilities with
respect to two distinct wave vectors occur resembles the situation
encountered in comb-coil diblock copolymers \cite{nakobrku01,nabr01}.
\section{The First Harmonics approximation}\label{sec_FHA}
In the previous section, we pointed out that the  
instability of the isotropic phase with respect to nematic fluctuations
gives rise to a global nematic ordering.
The density order parameters are constrained;
the volume integral of $\psi_C$ and $\psi_R$ should vanish.
Therefore, the Fourier components $\psi_C(p)$ and $\psi_R(p)$ can only give
contributions for nonzero $p$.
The nematic order parameter is not constrained, 
and it predominantly describes global ordering corresponding to the 
zero mode $p=0$ in Fourier space.

The Landau free energy given in Eq.~(\ref{lanentr}) is not solvable as
it is.  A way to solve the minimization condition for
Eq.~(\ref{lanentr}) is to expand the order parameters in harmonics
representing certain discrete symmetries.  Therefore, we shall adopt
the first harmonics approximation (FHA) for the density order
parameter $\psi=\psi_R=-\psi_C$, {\em i.e.}, we take in momentum space
\bea 
\psi(\vec k)\simeq\frac{\psi}{\sqrt{n_S}}\sum_{\vec Q\in H_S}
E_\phi(\vec Q) \delta_K(\vec Q-\vec k), \qquad |\vec
Q|=q^\ast,\label{anzord1} 
\eea 
with $q^\ast$ the radius of the first
harmonics sphere, and where $E_\phi(\vec Q)$ is a phase factor and
$H_S$ is the set of wave vectors $\vec Q$ with radius $|\vec
Q|=q^\ast$ describing a particular discrete symmetry $S$ of the
structure.  The number $n_S$ is half the number of vectors in $H_S$.
For the lamellar morphology $n_{lam}=1$, for hexagonal $n_{hex}=3$,
for bcc $n_{bcc}=6$.  Moreover, the phase factor $E_\phi(\vec Q)=1$
for the basic morphologies of interest.  Clearly, the FHA restricts
the microphase morphologies to morphologies which can be described by
a single amplitude or order parameter $\psi$.  Especially the nematic
ordering could give rise to more complex morphologies which require
more than one amplitude.  To keep the analysis manageable,
morphologies, which would require more than one amplitude, are not
considered in this paper.

As explained in the previous section, we can assume that
the nematic ordering is global and thus described by a space 
independent order parameter proportional to 
the nematic director $\eta^\mu\eta^\nu -\delta^{\mu\nu}/3$.
Thus the Ansatz for $S^{\mu\nu}$ is
\bea
S^{\mu\nu}(\vec k)&\simeq& S\delta_K(\vec k)
\left(\eta^\mu\eta^\nu-\frac{\delta^{\mu\nu}}{3}\right)
=S\delta_K(\vec k){\cal N}^{\mu\nu}
,\label{anzord2}\\
{\cal N}^{\mu\nu}&\equiv& \left(\eta^\mu\eta^\nu-\frac{\delta^{\mu\nu}}{3}\right),
\qquad {\cal N}^{\mu\nu}{\cal N}^{\mu\nu}
=\frac{2}{3}.\label{Nudef}
\eea
where $S$ on right-hand side is the space independent order parameter, 
and $|\vec \eta|=1$. 
With the Ans\"atze (\ref{anzord1}) and (\ref{anzord2}), 
the free energy can be expressed as
\bea
{\cal F}_S&=& \left[
c^{(2)}_{\psi\psi} -2N\chi\right]\psi^2+
\left[c^{(2)}_{SS}-\frac{N\omega}{3}\right] S^2+
c^{(3)}_{\psi\psi\psi}\psi^3+c^{(3)}_{\psi\psi S}\psi^2S
+c^{(3)}_{SSS}S^3\nonumber\\
&+&c^{(4)}_{\psi\psi\psi\psi} \psi^4
+c^{(4)}_{\psi\psi\psi S} \psi^3 S+c^{(4)}_{\psi\psi S S } \psi^2 S^2
+c^{(4)}_{SSSS} S^4,\label{freeenergy3}
\eea
where terms of the form 
$\psi S$, $\psi S^2$ and $\psi S^3$ are absent in this approximation.
The $c$ coefficients are given in terms of $\gamma$ functions,
which are related to the $\Gamma$ functions defined in Sec.~\ref{sec_lamf} 
(see below):
\bea
c^{(2)}_{\psi\psi}&=&\frac{1}{2n_S}\sum_{Q_1}\sum_{Q_2}
\delta_K(Q_1+Q_2)
\gamma^{(2)}_{\psi\psi}(Q_1),\\
c^{(2)}_{S S}&=&\frac{1}{2} \gamma^{(2)}_{S S}(0),\\
c^{(3)}_{\psi\psi\psi}&=&\frac{1}{3!n_S\sqrt{n_S}}
\sum_{Q_1}\sum_{Q_2}\sum_{Q_3}
\delta_K(Q_1+Q_2+Q_3)\gamma^{(3)}_{\psi\psi\psi}(Q_1,Q_2),\\
c^{(3)}_{\psi\psi S}&=&\frac{1}{2 n_S}\sum_{Q_1}\sum_{Q_2}
\delta_K(Q_1+Q_2)\gamma^{(3)}_{\psi\psi S}(Q_1,Q_2),\\
c^{(3)}_{S S S}&=&\frac{1}{3!}\gamma^{(3)}_{SSS}(0,0),
\eea
and
\bea
c^{(4)}_{\psi\psi\psi\psi}&=&\frac{1}{4!n_S^2}
\sum_{Q_1}\sum_{Q_2}\sum_{Q_3}\sum_{Q_4}
\delta_K(Q_1+Q_2+Q_3+Q_4)\gamma^{(4)}_{\psi\psi\psi\psi}(Q_1,Q_2,Q_3),\\
c^{(4)}_{\psi\psi\psi S}&=&\frac{1}{3!n_S\sqrt{n_S}}
\sum_{Q_1}\sum_{Q_2}\sum_{Q_3} \delta_K(Q_1+Q_2+Q_3)
\gamma^{(4)}_{\psi\psi\psi S}(Q_1,Q_2,Q_3),\\
c^{(4)}_{\psi\psi S S}&=&\frac{1}{4 n_S}
\sum_{Q_1}\sum_{Q_2} \delta_K(Q_1+Q_2)
\gamma^{(4)}_{\psi\psi S S}(Q_1,Q_2,0),\\
c^{(4)}_{S S S S}&=&\frac{1}{4!}\gamma^{(4)}_{S S S S}(0,0,0).
\eea
In the above formulas the $\gamma$ function are related to the $\Gamma$ 
functions of Eq.~(\ref{lanentr}) in the following way.
The second order $\gamma$ functions are
\bea
\gamma^{(2)}_{\psi\psi}(Q)&\equiv&\epsilon_a\epsilon_b \Gamma^{(2)}_{ab}(Q)
=\Gamma^{(2)}_{RR}(Q)+\Gamma^{(2)}_{CC}(Q)-2\Gamma^{(2)}_{RC}(Q)
,\\
\gamma^{(2)}_{S S}(0)&\equiv &{\cal N}^{\mu\nu}{\cal N}^{\rho\sigma}
\Gamma^{(2)\mu\nu\rho\sigma}_{SS}(0)
={\cal N}{\cal N}
\Gamma^{(2)}_{SS}(0),
\eea
with ${\cal N}$ given in Eq.~(\ref{Nudef}), and  
where $\psi_R=\epsilon_R \psi$, $\psi_C=\epsilon_C \psi$,
\bea
\epsilon_R=1,\qquad\epsilon_C=-1.
\eea
The third order $\gamma$ functions are (with $|Q_1|=|Q_2|=|Q_1+Q_2|$)
\bea
\gamma^{(3)}_{\psi\psi\psi}(Q_1,Q_2)&\equiv&
\epsilon_a\epsilon_b\epsilon_c \Gamma^{(3)}_{abc}(Q_1,Q_2),\\
\gamma^{(3)}_{\psi\psi S}(Q_1,-Q_1)&\equiv&
\epsilon_a\epsilon_b{\cal N}\Gamma^{(3)}_{abS}(Q_1,-Q_1),\\
\gamma^{(3)}_{SSS}(0,0)&\equiv&{\cal N}{\cal N}{\cal N}\Gamma^{(3)}_{SSS}(0,0).
\eea
Finally the fourth order $\gamma$ functions are
(with $|Q_1|=|Q_2|=|Q_3|=|Q_1+Q_2+Q_3|$)
\bea
\gamma^{(4)}_{\psi\psi\psi\psi}(Q_1,Q_2,Q_3)&\equiv&
\epsilon_a\epsilon_b\epsilon_c\epsilon_d \Gamma^{(4)}_{abcd}(Q_1,Q_2,Q_3),\\
\gamma^{(4)}_{\psi\psi\psi S}(Q_1,Q_2,-Q_1-Q_2)
&\equiv&\epsilon_a\epsilon_b\epsilon_c{\cal N} 
\Gamma^{(4)}_{abcS}(Q_1,Q_2,-Q_1-Q_2),\\
\gamma^{(4)}_{\psi\psi S S}(Q_1,-Q_1,0)
&\equiv&\epsilon_a\epsilon_b{\cal N} {\cal N} 
\Gamma^{(4)}_{abS S}(Q_1,-Q_1,0),\\
\gamma^{(4)}_{S S S S}(0,0,0)
&\equiv&{\cal N} {\cal N} {\cal N} {\cal N} 
\Gamma^{(4)}_{S S S S}(0,0,0).
\eea

The coefficients $c^{(2)}$, $c^{(3)}$, $c^{(4)}$ depend on $N$, $f_R$,
$q^\star$, $\omega/\chi$, $\chi$ and the symmetry-group (LAM, HEX,
BCC).  Moreover, the coefficients $c^{(3)}$ and $c^{(4)}$ also depend
on the angle $\theta$ between the director $\eta$ and the orientation
of the lattice of the microphase symmetry.  The characteristic wave
vector $q^\star$ depends on the architecture, $N$ and $f_R$.

\section{Minimization of the free energy}\label{sec_minFE}
Upon minimization of the free energy Eq.~(\ref{freeenergy3}),
with respect to the order parameters $\psi$ and $S$ and the angle $\theta$,
we shall distinguish between seven different phases.
These phase are:
\begin{itemize}
\item[1.]{The isotropic phase (I): $\psi=0$, $S=0$}
\item[]{Microphases (M): $\psi\not =0$, $S=0$
\begin{itemize}
\item[2.]{Lamellar (LAM)}
\item[3.]{Hexagonal (HEX)}
\item[4.]{BCC (BCC)}
\end{itemize}
}
\item[5.]{Nematic phase (N): $\psi=0$, $S\not =0$}
\item[]{Smectic phases (S): $\psi\not =0$, $S\not =0$}
\begin{itemize}
\item[6.]{Smectic A: $\theta=0$}
\item[7.]{Smectic C: $0<\theta\leq \pi/2$}
\end{itemize}
\end{itemize}
The smectic phases A and C are defined as lamellar density fluctuating 
microphases with either a nematic director parallel to the lamellar director (A) 
or with a nonzero angle $\theta$ between them (C).  

As mentioned previously we assume that temperature and architecture are
the two dominant parameters to play with, and therefore, 
following the reasoning by Singh {\em et al.} \cite{sigolifr94},
we assume that both $\chi$ and $\omega$ scale inversely with the temperature $T$. 
This means that for a fixed ratio $r=\omega/\chi$ and fixed $N$, 
we can draw the usual two-dimensional $\chi N$ vs $f$ ``phase diagrams''. 
 
Clearly, in this fourth order Landau expansion the order parameters $\psi$ and $S$
should be reasonably small, {\em e.g.}, $\psi,\, S\ll 1$.
For the usual microphase separation this is the case close to the critical point. 
However, the isotropic-nematic transition is first order due to the presence of
the third order term in $S$ in Eq.~(\ref{freeenergy3}). 
Therefore, we investigate first the ``weakness'' of the first order
isotropic-nematic transition.
Assuming that we are in the region of parameter space such that $\psi=0$, 
we are left with the free energy
\bea
{\cal F}^{nem}_S&=&
\left[c^{(2)}_{SS}-\frac{N\omega}{3}\right] S^2
+c^{(3)}_{SSS}S^3
+c^{(4)}_{SSSS} S^4,\label{FES}
\eea
where the coefficients are explicitly given in the appendix.
The coefficients are
\bea
c^{(2)}_{SS}=\frac{5}{2 f_R^2},
\quad c^{(3)}_{SSS}=-\frac{25}{21 f_R^3},\quad
c^{(4)}_{SSSS}=\frac{425}{196 f_R^4}.
\eea
The free energy (\ref{FES}) has a first order phase transition to a nematic phase when
\bea
\left[c^{(2)}_{SS}-\frac{N\omega}{3}\right]\leq 
\frac{[c^{(3)}_{SSS}]^2}{4 c^{(4)}_{SSSS}}\quad \Longrightarrow \quad
N\omega_c=3\left\{c^{(2)}_{SS}-\frac{[c^{(3)}_{SSS}]^2}{4 c^{(4)}_{SSSS}}\right\}
=\frac{1}{f_R^2}\left(\frac{715}{102}\right),
\eea
where $\omega_c$ is the binodal value.\footnote{The value 
$\omega_s=15/(2 f_R^2)$ is the so-called spinodal value corresponding 
to the change of sign of the second order term.}
The critical value of $S$ is given by
\bea
S_c=-\frac{c^{(3)}_{SSS}}{2 c^{(4)}_{SSSS}}=\frac{14}{51}f_R\leq \frac{14}{51}\approx 0.27. \label{Scdef}
\eea
This is a typical value for $S_c$ for nematic liquids \cite{chlu}.
Thus for large rod fractions $f_R$, the nematic order parameter is close to $1/3$.
It is well established that the Landau expansion for coil-coil AB diblocks
is highly weakly first order even far from the critical point in the
$N\chi$-$f$ plane. There, the critical value, $\psi_c$, for the order parameter 
$\psi$ is at least one order of magnitude less than $S_c$ 
({\em e.g.}, $\psi_c\simeq 1/30$).

Now, suppose that we are in a regime (coil rich) of parameter space where 
the nematic order parameter is zero ($S=0$) and consider the minimization
of Eq.~(\ref{freeenergy3}) with respect to $\psi$.
The free energy in this region reads
\bea
{\cal F}_S&=& \left[
c^{(2)}_{\psi\psi}-2N\chi\right]\psi^2+c^{(3)}_{\psi\psi\psi}\psi^3
+c^{(4)}_{\psi\psi\psi\psi} \psi^4.\label{FEM}
\eea
The $q^\ast$ was already (and can still be) determined from the
minimum of $c^{(2)}_{\psi\psi}$ with respect to $|Q|$.
By definition this minimum is $q^\ast$.
The spinodal is then given by the equation
\bea
2N\chi_s=c^{(2)}_{\psi\psi}.
\eea
 
In the FHA the free energy for a lamellar morphology corresponding to $q^\ast$ 
is known \cite{le80} to be given by
\bea
{\cal F}^{lam}_S&=& \left[
c^{(2)lam}_{\psi\psi}-2N\chi\right]\psi_l^2
+c^{(4)lam}_{\psi\psi\psi\psi} \psi_l^4,\label{FElam}
\eea 
where (with $|Q|=q^\ast$)
\bea
c^{(2)lam}_{\psi\psi}&=&\frac{1}{2}\sum_{Q_1\in lam}
\sum_{Q_2\in lam}\delta_K(Q_1+Q_2)
\gamma^{(2)}_{\psi\psi}(Q_1)=\gamma^{(2)}_{\psi\psi},\\
c^{(4)lam}_{\psi\psi\psi\psi}&=&\frac{1}{4}\gamma^{(4)}_{\psi\psi\psi\psi 1},
\eea
with $\gamma^{(2)}_{\psi\psi}$ and $\gamma^{(4)}_{\psi\psi\psi\psi 1}$
explicitly given in the appendix.
The  hexagonal and bcc morphologies
have free energies:
\bea
{\cal F}^{hex}_S&=& \left[
c^{(2)hex}_{\psi\psi}-2N\chi\right]\psi_h^2
+c^{(3)hex}_{\psi\psi\psi}\psi_h^3
+c^{(4)hex}_{\psi\psi\psi\psi} \psi_h^4,\label{FEhex} \\
{\cal F}^{bcc}_S&=& \left[
c^{(2)bcc}_{\psi\psi}-2N\chi\right]\psi_b^2
+c^{(3)bcc}_{\psi\psi\psi}\psi_b^3
+c^{(4)bcc}_{\psi\psi\psi\psi} \psi_b^4,\label{FEbcc} 
\eea
with
\bea
c^{(2)hex}_{\psi\psi}=\gamma^{(2)}_{\psi\psi},\quad
c^{(3)hex}_{\psi\psi\psi}=\frac{2}{3\sqrt{3}}\gamma^{(3)}_{\psi\psi\psi},\quad
c^{(4)hex}_{\psi\psi\psi\psi}=\frac{1}{12}\gamma^{(4)}_{\psi\psi\psi\psi 1}+
\frac{1}{3}\gamma^{(4)}_{\psi\psi\psi\psi 2},
\eea
respectively,
\bea
c^{(2)bcc}_{\psi\psi}&=&\gamma^{(2)}_{\psi\psi},\quad
c^{(3)bcc}_{\psi\psi\psi}=\frac{4}{3\sqrt{6}}\gamma^{(3)}_{\psi\psi\psi},\nonumber\\
c^{(4)bcc}_{\psi\psi\psi\psi}&=&
\frac{1}{24}\gamma^{(4)}_{\psi\psi\psi\psi1}
+\frac{1}{3}\gamma^{(4)}_{\psi\psi\psi\psi2}
+\frac{1}{12}\gamma^{(4)}_{\psi\psi\psi\psi3}
+\frac{1}{6}\gamma^{(4)}_{\psi\psi\psi\psi4}.
\eea
For fixed $N \chi$, $r$, $f_R$ and $N$, it is then straightforward to find the 
morphology with lowest free energy.

The next step is to consider the region of the phase diagram where
$\psi \ll S <1$. In this case if the temperature is lowered first the
isotropic to nematic transition is encountered meaning that the
lowering of the free energy is primarily driven by the appearance of a
nonzero value for $S$. When the temperature is decreased ($N\chi
\uparrow$), the appearance of a microphase structure ($\psi\not =0$)
lowers the free energy further.  In this particular case, the value of
$S$ is mainly determined by Eq.~(\ref{FES}).  The value of $\psi$ then
follows from the free energy part of Eq.~(\ref{freeenergy3}), 
\bea
{\cal F}^{mix}_S&=& \left[c^{(2)}_{\psi\psi}+c^{(3)}_{\psi\psi S} S
  +c^{(4)}_{\psi\psi S S }S^2-2N\chi\right]\psi^2
+\left[c^{(3)}_{\psi\psi\psi}+c^{(4)}_{\psi\psi\psi S}S\right]\psi^3
+c^{(4)}_{\psi\psi\psi\psi} \psi^4,\label{FEmix} 
\eea 
in which $S$ should be considered as an external field.  The total free energy
being ${\cal F}_S={\cal F}^{nem}_S+{\cal F}^{mix}_S$.  In the above
Eq.~(\ref{FEmix}) the angle $\theta$ comes into play.

For instance, if the morphology is assumed to be lamellar, 
the coupling term $c^{(3)}_{\psi\psi S}$ can be written as
\bea
c^{(3)lam}_{\psi\psi S}&=&\frac{\gamma^{(3)}_{\psi\psi S}}{2 n_{lam}}
\sum_{Q\in lam} P_2(\hat Q\cdot \eta)=
\frac{\gamma^{(3)}_{\psi\psi S}}{2}
\left[P_2(\hat Q\cdot \eta)+P_2(-\hat Q\cdot \eta)\right]\nonumber\\
&=&P_2(\cos\theta)\gamma^{(3)}_{\psi\psi S},
\eea
with $\hat Q=\vec Q/q^\ast$, $\cos\theta=\hat Q\cdot \eta$.
In case of a hexagonal morphology (with six lattice vectors $\vec Q_i$),
we get
\bea
c^{(3)hex}_{\psi\psi S}&=&\frac{\gamma^{(3)}_{\psi\psi S}}{2 n_{hex}}
\sum_{Q\in hex} P_2(\hat Q\cdot \eta)
=\frac{\gamma^{(3)}_{\psi\psi S}}{3}
\sum_{i=1}^3 P_2(\hat Q_i\cdot \eta),\label{cppshex}
\eea
and for the bcc morphology
\bea
c^{(3)bcc}_{\psi\psi S}&=&\frac{\gamma^{(3)}_{\psi\psi S}}{2 n_{bcc}}
\sum_{Q\in bcc} P_2(\hat Q\cdot \eta)
=\frac{\gamma^{(3)}_{\psi\psi S}}{6}
\sum_{i=1}^6 P_2(\hat Q_i\cdot \eta),\label{cppsbcc}
\eea
The other coupling terms in Eq.~(\ref{FEmix}) are decomposed as
\bea
c^{(4)sym}_{\psi\psi SS}&=&\frac{1}{4 n_{sym}}\sum_{Q\in sym}
\left[
\gamma^{(4)}_{\psi\psi S S 0}
+\gamma^{(4)}_{\psi\psi S S 1} P_2(\hat Q\cdot \eta)
+\gamma^{(4)}_{\psi\psi S S 2} P_2^2(\hat Q\cdot \eta)
\right]\nonumber\\
&=&
\frac{1}{2}\left[\gamma^{(4)}_{\psi\psi S S 0}
+\gamma^{(4)}_{\psi\psi S S 1} A_{1 sym}(\Omega)
+\gamma^{(4)}_{\psi\psi S S 2} A_{2 sym}(\Omega)
\right], \label{cppss}
\eea
where $sym$ stands for either one of three basic morphologies,
and
\bea
A_{1 sym}(\Omega)=\frac{1}{2 n_{sym}}\sum_{Q\in sym} P_2(\hat Q\cdot \eta),
\qquad
A_{2 sym}(\Omega)=\frac{1}{2 n_{sym}}\sum_{Q\in sym} P_2^2(\hat Q\cdot \eta),
\eea
with $\Omega$ the three dimensional space angle between the nematic vector
$\vec\eta$ and the orientation vector of the lattice corresponding to the
space symmetry group $sym$.

Subsequently,
the free energy is written as
\bea
{\cal F}^{mix}_S&=&\alpha_1 \psi^2
-\alpha_2\psi^3
+\alpha_3\psi^4,\label{FEmix2}
\eea
with 
\bea
\alpha_1&=&c^{(2)}_{\psi\psi}+c^{(3)}_{\psi\psi S} S
+c^{(4)}_{\psi\psi S S }S^2-2N\chi,\\
\alpha_2&=&-\left[c^{(3)}_{\psi\psi\psi}+c^{(4)}_{\psi\psi\psi S}S\right],\\
\alpha_3&=&c^{(4)}_{\psi\psi\psi\psi}.
\eea
Then for the lamellar morphology (thus smectic), 
these coefficients can be expressed as
\bea
\alpha_1&=&-2N\chi+\beta_0(q^\ast)+\beta_1(q^\ast)P_2(\cos\theta)
+\beta_2(q^\ast)P_2^2(\cos\theta),\nonumber\\
\alpha_2&=&0,\qquad \alpha_3=\frac{1}{4}\gamma^{(4)}_{\psi\psi\psi\psi1},
\eea
where $\cos\theta=(\eta\cdot \vec Q)/q^\ast$.
The functions $\beta_i$ are
\bea
\beta_0=c^{(2)}_{\psi\psi}+\frac{1}{2}\gamma^{(4)}_{\psi\psi S S 0} S^2,
\qquad
\beta_1=\gamma^{(3)}_{\psi\psi S} S+\frac{1}{2}\gamma^{(4)}_{\psi\psi S S 1} S^2,
\qquad
\beta_2=\frac{1}{2}\gamma^{(4)}_{\psi\psi S S 2} S^2.
\eea
Now the minimum of ${\cal F}^{mix}_S$ with respect to $\theta$
corresponds to the minimum of $\alpha_1$, since $\alpha_3$ does 
not depend on $\theta$.
Since $\beta_1<0$ and $\beta_2>0$, the minimum of $\alpha_1$ with respect 
to $\cos\theta$ is
\bea
\alpha_1=-2N\chi+\beta_0-\frac{\beta_1^2}{4\beta_2},
\eea
with
\bea
\theta_{min}=\arccos\left(\sqrt{\frac{\beta_2-\beta_1}{3\beta_2}}\right).
\eea
Numerically it turns out that $\gamma^{(4)}_{\psi\psi SS2}$ is roughly
one order of magnitude larger than $\gamma^{(4)}_{\psi\psi SS1}$ and 
$\gamma^{(4)}_{\psi\psi SS0}$ for a large range of volume fractions $f_C$.
Since the nematic order parameter $S$ is close to $1/3$ or even larger 
({\em e.g.}, see Eq.~(\ref{Scdef})), 
the consequence is that in general $\beta_2\gg \beta_1$ and therefore the
angle $\theta$ is close to 
\bea
\theta_m\approx \arccos\sqrt{1/3}\approx 54^o.
\eea
Thus a smectic C phase with an angle of about $54^o$ is obtained
in the region of parameter space where nematic instabilities 
precede density fluctuations.

However, the above conclusion is premature unless the other two 
morphologies are considered.
For these morphologies, besides the terms already given in 
Eqs.~(\ref{cppshex}), 
(\ref{cppsbcc}), and (\ref{cppss}), 
third order density terms are present and consequently 
the coupling term 
\bea
c^{(4)}_{\psi\psi\psi S sym}=\frac{1}{3!n_{sym}\sqrt{n_{sym}}}
\sum_{Q_1\in sym}\sum_{Q_2\in sym}\gamma^{(4)}_{\psi\psi\psi S}(Q_1,Q_2,-Q_1-Q_2),
\eea
plays a role.
This term 
is written as
\bea
c^{(4)}_{\psi\psi\psi S hex}&=&\frac{2}{3\sqrt{3}}
\sigma_{\psi\psi\psi S}
\left[P_2(\hat Q_1 \cdot \eta)
+P_2(\hat Q_2 \cdot \eta)
+P_2(\hat Q_3 \cdot \eta)
\right],\\
c^{(4)}_{\psi\psi\psi S bcc}&=&\frac{4}{3\sqrt{6}}
\sigma_{\psi\psi\psi S}
\left[P_2(\hat Q_1 \cdot \eta)
+P_2(\hat Q_2 \cdot \eta)
+P_2(\hat Q_3 \cdot \eta)
\right],
\eea
with $Q_3=-Q_1-Q_2$ and $\sigma_{\psi\psi\psi S}$ defined 
in Eq.~(\ref{sigpppsdef}).
The three vectors $Q_i$ can be an arbitrary triangle in the vector space
of the first harmonic sphere of the particular lattice symmetry group (HEX or BCC).
However, by comparing numerically the free energies of the hexagonal and bcc 
morphology with nonzero $S$ in the FHA, it turns out that
the smectic C phase 
(thus lamellar) has the lowest free energy 
in this particular region of the parameter space.

In the coil rich region of the parameter space, it is possible to take a value
for $r=\omega/\chi$ so that microphase separation 
occurs prior to nematic ordering, {\em i.e.}, with $S\ll\psi<1$, 
on decreasing the temperature.
In this case the order parameter $\psi$ is mainly determined by Eq.~(\ref{FEM}).
However, now the nematic order parameter $S$ is driven by the field $\psi$ 
which acts as an external field in the free energy part
\bea
{\cal F}_{mix}&=& \left[c^{(3)}_{\psi\psi S}\psi^2+
c^{(4)}_{\psi\psi\psi S} \psi^3\right] S 
+\left[c^{(2)}_{SS}+c^{(4)}_{\psi\psi S S } \psi^2-\frac{N\omega}{3}\right] S^2,\label{FESmix}
\eea
up to second order in $S$.
We assume that the temperature is still above the binodal temperature
(thus below $N\chi_c$) so that nematic ordering is purely density
driven.
Then
\bea
S_m\simeq  -\frac{\left[c^{(3)}_{\psi\psi S}\psi^2+
c^{(4)}_{\psi\psi\psi S} \psi^3\right]}{2\left[
c^{(2)}_{SS}-N\omega/3\right]}.
\eea
For a large region of the phase diagram with the above 
constraint $S\ll\psi<1$ it turns out that $S_m\ll S_c$ as given by
Eq.~(\ref{Scdef}).
This concludes the analysis of the free energy, Eq.~(\ref{freeenergy3}).

\section{The phase diagram}\label{sec_PD}
\begin{figure}
\includegraphics{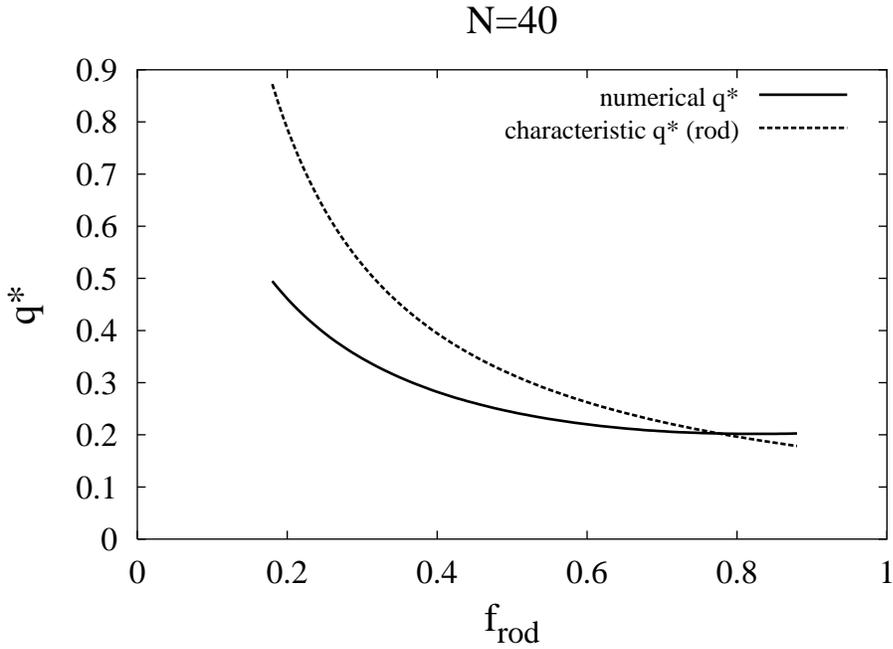}
\caption{The numerical value of the wave vector $q^\ast$ for $N=40$ 
as minimum of $\gamma^{(2)}_{\psi\psi}$ vs 
the characteristic rod wave vector 
$q^\ast=2\pi/(N f_R)$.}
\label{fig_qstar}
\end{figure}

In the previous section, we outlined the various steps in the minimization of
the FHA free energy Eq.~(\ref{freeenergy3}).
This section is devoted to the actual computation and derivation of
the phase diagrams.
The first step in the process is to determine $q^\ast$.
This wave vector $q^\ast$ is given by the minimum of $\gamma^{(2)}_{\psi\psi}(q)$
of Eq.~(\ref{g2ppdef}) with respect to $q$.
It is to be expected that the characteristic length scale 
for microphase separation is predominantly determined by the length $\ell =Nf_R$
of the rod part of the diblock. At least for large $N$ and in the 
rod rich region this is to be expected, since the length scale
of the rod part scales with $N$ and that of the coil part with $N^{1/2}$.
This means that the lowest $q$ vector for the rod is much smaller than the 
lowest $q$ vector of the coil part. 
The corresponding characteristic 
wave vector for the rod is $q^\ast=2\pi/\ell=2\pi/(Nf_R)$. 
In Fig.~\ref{fig_qstar} the numerical value of $q^\ast$ is depicted vs 
rod fraction $f_R$ and compared with the above-mentioned characteristic 
wave vector. From this figure it is clear that for a large region of $f_R$
both $q^\ast$ lie rather close together, supporting the view that the rod-length
scale characterizes the microphases.

With $q^\ast$ numerically computed as function of $f_R$, 
the higher point vertices $\gamma^{(3)}$ and $\gamma^{(4)}$ 
are determined, since these do not depend on $\chi$ and $\omega$.
Subsequently all $c$ coefficients of 
the Landau free energy Eq.~(\ref{freeenergy3}) are computed.
Depending on the values of $\chi$, $r$, $N$ the phase diagram is constructed
by minimizing the free energy with respect to $\psi$, $S$, and $\theta$.

\begin{figure}
\includegraphics{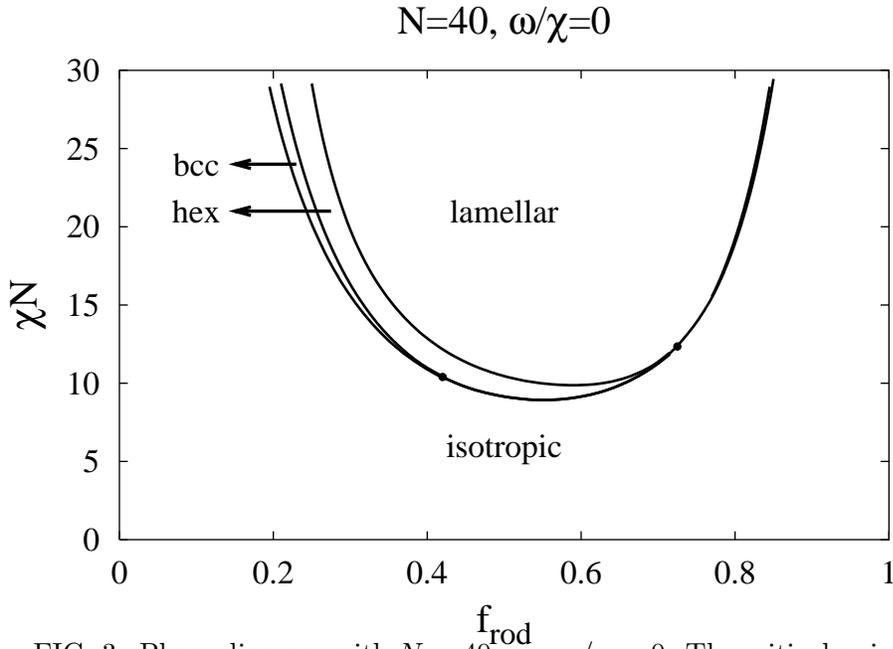}
\caption{Phase diagram with $N=40$, $r=\omega/\chi=0$. The critical point
lies in the rod-rich region near $f_{rod}\simeq 0.73$
and there is a triple point near $f_{rod}\simeq 0.42$,
separating the isotropic/BCC/HEX phases.}
\label{fig_pdn40r0}
\end{figure}

As pointed out in the previous section, 
when there is no Maier-Saupe interaction, 
there will be no accountable or notable value for the
nematic order parameter $S$ ({\em i.e.}, $S\ll 1/3$).
The phase diagram for this situation where $\omega=0$, {\em i.e.}, $r=0$, 
is given in Figure~\ref{fig_pdn40r0} for $N=40$. 
The asymmetry of the phase diagram with respect to the AB coil-coil 
diblock phase diagram (as first derived by Leibler~\cite{le80}) is apparent.
First of all, the critical point, which is given by the root of 
the $\gamma^{(3)}_{\psi\psi\psi}$ vertex,
is shifted to the rod rich part.
Beyond the critical point for even higher rod fraction the bcc 
phase is absent.                                              

The coil rich part of the phase diagram is quite similar to an AB
coil-coil diblock phase diagram with first the appearance of a bcc
phase prior to hexagonal and lamellar phases.  However, close to
fraction $f_R\sim 0.4$ the bcc phase disappears at the triple point
and for increasing $\chi$ only the hexagonal and lamellar phases are
encountered.  This is a result of the relative smallness of 
the ratio $k$,
\begin{eqnarray}
k\equiv 
\left(\frac{c^{(3)bcc}_{\psi\psi\psi}}{c^{(3)hex}_{\psi\psi\psi}}\right)^2 
\frac{c^{(4)hex}_{\psi\psi\psi\psi}}
{c^{(4)bcc}_{\psi\psi\psi\psi}},
\end{eqnarray}
as compared to coil-coil diblock copolymers \cite{le80}.
Whenever $k<1$ the hexagonal phase is encountered first upon crossing
the binodal curve from the isotropic phase.  The triple point
corresponds to $k=1$.

\begin{figure}
\includegraphics{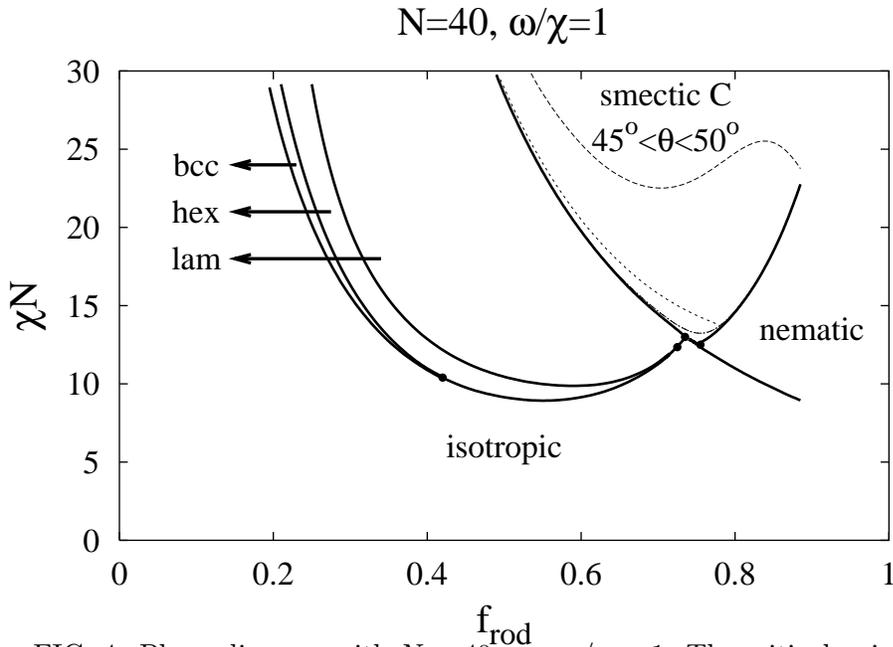}
\caption{Phase diagram with $N=40$, $r=\omega/\chi=1$.
The critical point still lies near $f_{rod}\simeq 0.73$.
The three other dots are triple points. The dashed curves in the smectic C phase 
are contour lines for the angle $\theta$, separating intervals of $5^o$. 
The smallest value for $\theta$ is about $30^o$ just at the phase boundaries 
near the triple points.}
\label{fig_pdn40r1}
\end{figure}

\begin{figure}
\includegraphics{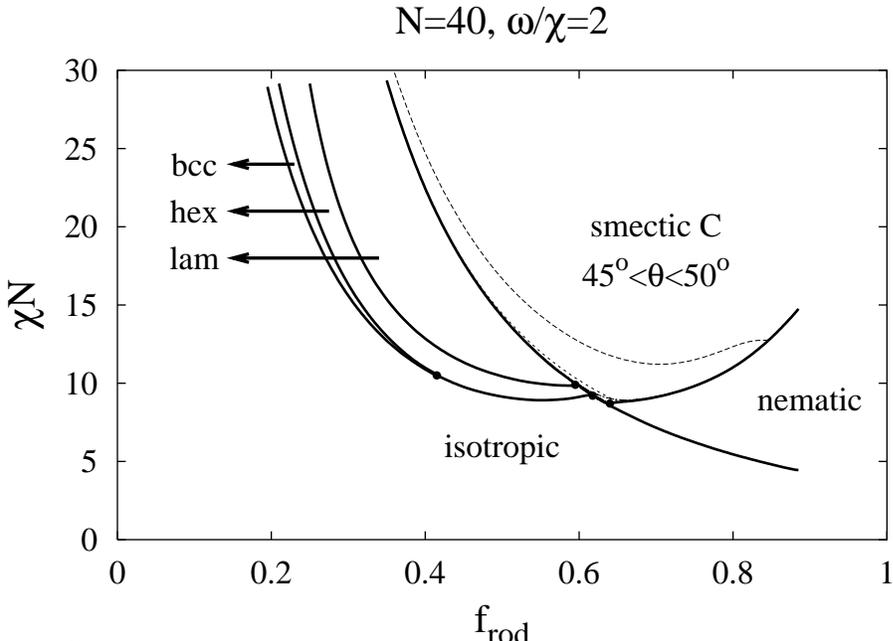}
\caption{Phase diagram with $N=40$, $r=\omega/\chi=2$.
There is no critical point. All dots are triple points. 
The dashed curves in the smectic C phase 
are contour lines for the angle $\theta$, separating intervals of $5^o$.}
\label{fig_pdn40r2}
\end{figure}

\begin{figure}
\includegraphics{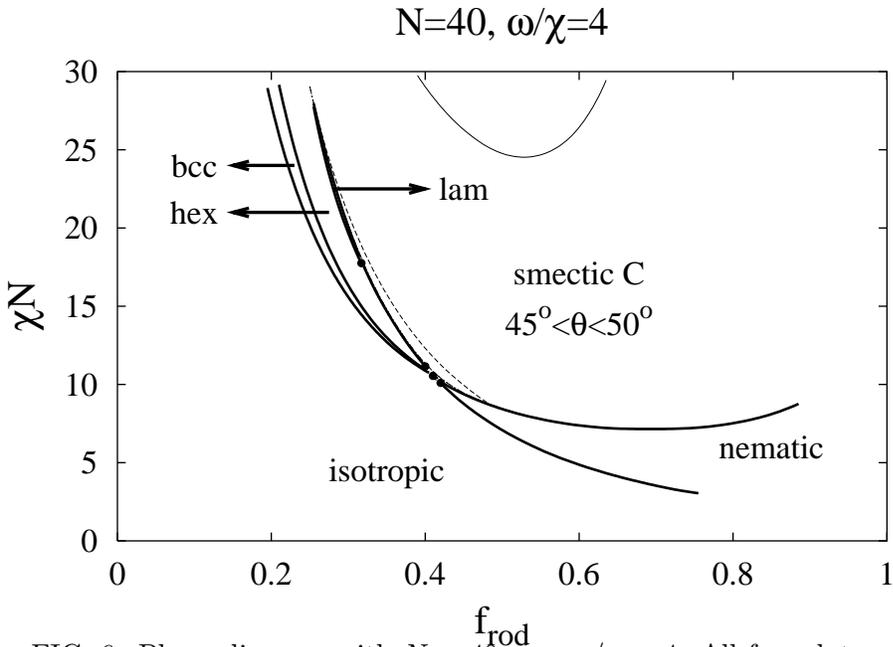}
\caption{Phase diagram with $N=40$, $r=\omega/\chi=4$. 
All four dots represent triple points.
The dashed curves and the thin solid curve in the smectic C phase 
are contour lines for the angle $\theta$, separating 
intervals of $5^o$.
}
\label{fig_pdn40r4}
\end{figure}

\begin{figure}
\includegraphics{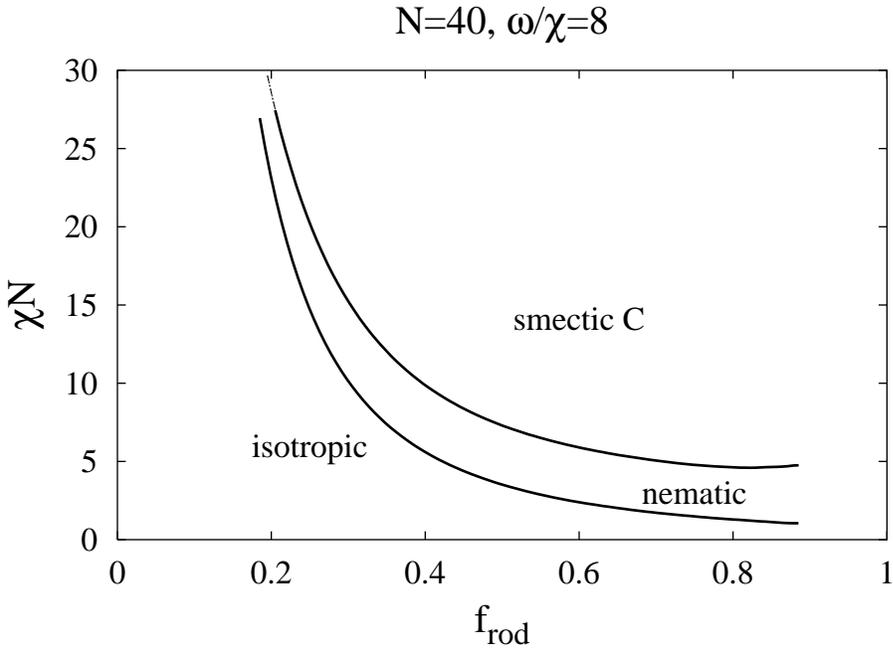}
\caption{Phase diagram with $N=40$, $r=\omega/\chi=8$. There are no triple points.}
\label{fig_pdn40r8}
\end{figure}

The dependence on the ratio $r$ is depicted in a series of figures,
Fig.~\ref{fig_pdn40r0}-Fig.~\ref{fig_pdn40r8}.
The isotropic-nematic phase boundary (PB) is 
roughly given by the spinodal curve
\bea
\chi_s=15/(2 r f_R^2),
\eea
see footnote in Sec.~\ref{sec_minFE}.
Thus by increasing $r$, $\chi_s$ is lowered which appears as a shifting of the 
isotropic-nematic PB to the left.
Whenever the nematic phase region ``overlaps'' with the microphase region
a smectic C phase (lamellar) is favored over the bcc and hexagonal 
morphologies; nematic ordering favors lamellae.
This results from the fact that, whenever a nonzero value for $S$ appears,
the second order coefficient $\alpha_1$ in the Landau free energy 
for $\psi$ (Eq.~(\ref{FEmix2})) 
is considerable smaller for the lamellar phase 
than for the hexagonal phase, after minimization with respect to angles.
Consequently, in the nematic phase region, 
the spinodal curve for the lamellar morphology (=smectic) 
lies below the binodal curves for the hexagonal and bcc morphologies.
In the FHA the hexagonal and bcc morphologies are incompatible with
nematic order, and a smectic (layered liquid crystalline) phase prevails.

Contour lines of the angle $\theta$ are depicted also 
in the figures. The dots denote ``triple points''.
The angle increases rapidly from $35$ to $40^o$ 
close to the nematic smectic-C phase boundary
to values between $45$ and $55^o$. This behavior is in agreement with
the analysis in the previous section.
In Fig.~\ref{fig_pdn40r8} the nematic phase region has completely 
overwhelmed the microphase region. Even in the coil rich region
there is an isotropic-nematic transition.
This means that the microphases are absent all together and only the 
smectic C phase appears for lower temperatures.

\begin{figure}
\includegraphics{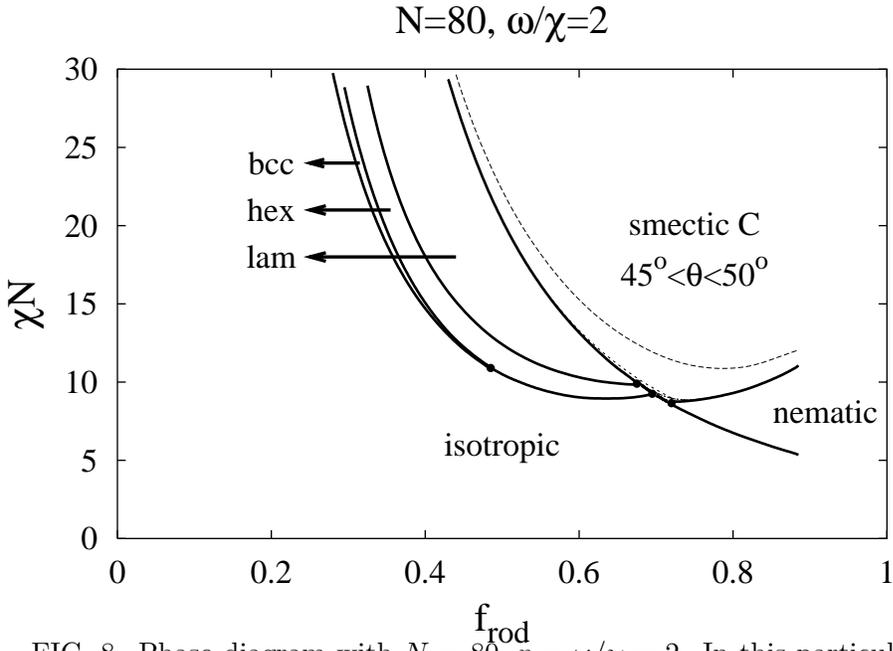}
\caption{Phase diagram with $N=80$, $r=\omega/\chi=2$.
In this particular case there is no critical point 
and the dots are triple points.
The dashed curves in the smectic C phase 
are contour lines for the angle $\theta$, separating 
intervals of $5^o$.
}
\label{fig_pdn80r2}
\end{figure}

If the overall length of the diblock is increased ($N\uparrow$) 
the critical point and the triple points will shift 
to the rod-rich part. However, qualitatively, the phase diagrams for 
fixed $r$ but with different $N$ are quite similar, as can be seen by
comparing Figs.~(\ref{fig_pdn40r2}) and (\ref{fig_pdn80r2}).

\section{Concluding remarks}\label{sec_concl}
In this paper, we studied an incompressible melt of rod-coil diblock
copolymers.  Both microphase separation and orientational ordering
were investigated.  We have derived a Landau expansion of the free
energy of the melt up to fourth order in the two order parameters
$\psi$ and $S$, representing, respectively, compositional and nematic
ordering.  The compositional ordering or microphase separation was
driven by the usual Flory-Huggins interaction, whereas the
orientational ordering was driven by a Maier-Saupe interaction.  Up to
seven different phases of the melt could be distinguished as function
of molecular architecture, temperature, and relative strength of the
molecular interactions. The characteristic length scale for the
microphases turned out to be roughly the length of the rod part of the
diblock.

In comparison with the coil-coil (AB) diblocks, the phase diagram for
rod-coil diblocks is quite asymmetric, with a critical point lying in
the rod-rich region on the spinodal curve.  Nevertheless, the spinodal
curve for the rod-coil system is nearly symmetric.  There is a
complete absence of the spherical or bcc microphase in the rod-rich
region, beyond the critical point.  This is perhaps not so surprising,
since it is hard to imagine spherical micelles of coil segments being
embedded in a matrix of rods.  The coil-rich region is more similar to
a diblock, where the spherical microphase appears prior to hexagonal
and lamellar structures when the temperature decreases.  However, for
intermediate rod-volume fractions, there is an interesting suppression
of the bcc morphologies with respect to the hexagonal one.  When
crossing the binodal curve from the isotropic phase, the hexagonal
phase appears instead of the bcc morphology.  There is an
isotropic-hexagonal-lamellar transition.  Such a transition is rather
uncommon for coil-coil (AB) diblocks.  The general features of our
phase diagram are in qualitative agreement with experiments on
rod-coil systems with small and intermediate rod volume
fractions \cite{rast94,racast97,lechohzi01,lekihwkiohzi01,sczayafith00}.
Moreover, in these experimental papers,
observations of bicontinuous cubic phases are reported.  The
description of these more complex phases is beyond the scope of the
present paper, since it would require the introduction of an
additional order parameter, namely, the amplitude of the second
harmonics.

Most of the features mentioned above are governed by compositional
ordering, which is most prominent in the coil-rich phase.  However,
different phase behavior is obtained when the orientational or nematic
ordering is the driving force in the melt.  Therefore, in the
rod-region, with reasonable values for the Maier-Saupe interactions
the main results are; the suppression of hexagonal and bcc structures
in the rod-rich region of the phase diagram whenever nematic ordering
occurs and the appearance of a smectic C phase in the rod-rich region,
with a characteristic angle $45^o\leq \theta<55^o$.

We have shown that when nematic phases appear prior to microphases
upon decreasing the temperature, instead of the usual bcc or hexagonal
morphologies a smectic C (lamellar) phase is obtained.  Also we have
found no evidence for smectic A phases in the rod rich region of the
phase diagram.  As was mentioned in the introduction, smectic C phases
and equivalent phases have been reported for experiments on rod-coil
systems for intermediate to high rod volume fractions 
\cite{chthobma96,muobth97}.
Moreover, the smectic angle $\theta$ is in quantitative agreement with
the experimental observed values for $\theta$, with $\theta\sim 45^o$.

At a first glance, the absence of a smectic A phase in the rod-rich
region might seem to contradict earlier theoretical results
\cite{maba98,seva86,ha89,sesu92}, where the orientational alignment is
chosen to be perfect (all rod segments are parallel, {\em i.e.},
$S\simeq 1$).  However, one should realize that the present model
describes a different limiting case of the rod-coil diblock copolymer
melt.  Clearly the free-energy expansion in order parameters breaks
down, whenever one of the order parameters becomes too large, {\em
  e.g.}, when $S$ approaches unity.  Roughly speaking, the
applicability of the Landau free energy is limited to the region in
the $\chi N-f_R$ plane, where the spinodal curves for order parameters
$\psi$ and $S$ intersect or lie reasonable close to one another.
Although, it can be shown that the Maier-Saupe interaction is a valid
approximation for the steric repulsion even when the alignment of
rod-segments is nearly or completely parallel \cite{hoos01}, the present
model cannot really be compared to the other works
\cite{maba98,seva86,ha89,sesu92} in great detail, due to the breakdown
of the Landau expansion for the order parameter $S$.

\acknowledgments{We would like to thank H.~Angerman, I.~Erukhimovich, 
S.~Kuchanov, R.~Nap, H. Slot, 
A.~Subbotin, and A.~Zvelindovsky for useful suggestions 
and stimulating discussions.
This research was supported by SOFTLINK, a technology-related 
soft condensed matter research program of the Dutch organization 
for scientific research NWO.}

\appendix
\section{The second order vertex 
{\protect\boldmath$\Gamma^{(2)}$}}\label{ap_gamma2}
In this appendix, we compute the second order vertex $\Gamma^{(2)}$.
The vertex $\Gamma^{(2)}$ is the inverse of the single chain correlation
function $W^{(2)}$ as given by Eq.~(\ref{gam2abdef}),
\bea
\Gamma^{(2)}_{\alpha\beta}(k_1)&=&
\left[W^{(2)}_{\alpha\beta}(k_1)\right]^{-1},
\eea
where the Greek index $\alpha=R,C,S$.
We introduce the matrix
\bea
{\bf W}^{\mu\nu\rho\sigma}\equiv \left(\begin{array}{ccc}
W^{(2)}_{RR}& W^{(2)}_{RC}& W^{(2)\rho\sigma}_{RS}\\
W^{(2)}_{CR}& W^{(2)}_{CC}& W^{(2)\rho\sigma}_{CS} \\
W^{(2)\mu\nu}_{SR} & W^{(2)\mu\nu}_{SC} & W^{(2)\mu\nu\rho\sigma}_{SS}
\end{array}\right),
\label{Wmatdef}
\eea
where
\bea
W^{(2)}_{RR}(p)&=&N^{-2}
\langle\hat\rho_R(\vec p)\hat\rho_R(-\vec p)\rangle_0=
\left(\frac{f_R}{\ell}\right)^2\int_0^{\ell}\int_0^{\ell}
ds ds^\prime\,
\langle e^{i(s-s^\prime)\vec p\cdot\vec u}\rangle_0
\nonumber\\
&=&f_R^2 K^{(2)}_{R1}(y)=g_{RR}(p),
\eea
with $f_R=1-f_C$, and
where the variable $y$ is
\be
y=N_R p.
\ee
The other functions are
\bea
W^{(2)}_{CC}(p)&=&f_C^2 K^{(2)}_{C1}(x)
=g_{CC}(p),\\
W^{(2)}_{RC}(p)&=&W^{(2)}_{CR}(p)
=f_C f_R K^{(1)}_C(x) K^{(1)}_R(y)=g_{RC}(p),
\eea
where the variable $x$ is
\bea
x=N_c p^2/6,
\eea
and $K^{(2)}_{C1}=f_D$ is the Debye function and the $K$ functions 
are defined in Appendix~\ref{ap_coilrodfies}.
Furthermore, the two point single chain functions with one 
nematic tensor is
\bea
W^{(2)\mu\nu}_{RS}(p)&=&W^{(2)\mu\nu}_{SR}(p)=
\left(\frac{p^\mu p^\nu}{p^2}-\frac{\delta^{\mu\nu}}{3}\right)
f_R^2 K^{(2)}_{RS}(y)=\Delta^{\mu\nu}g_{RS}(p),\\
W^{(2)\mu\nu}_{CS}(p)&=&W^{(2)\mu\nu}_{SC}(p)
=\left(\frac{p^\mu p^\nu}{p^2}-\frac{\delta^{\mu\nu}}{3}\right)
f_R f_C
K^{(1)}_C(x) K^{(1)}_{S0}(y)=\Delta^{\mu\nu}g_{CS}(p),
\eea
with $\Delta^{\mu\nu}=p^\mu p^\nu/p^2-\delta^{\mu\nu}/3$.
Finally, the two point function with two nematic tensors is expressed as 
\bea
W^{(2)\mu\nu\rho\sigma}_{SS}(p)&=&
f_R^2\sum_{i=1}^3 K^{(2)}_{Si}(y){\cal T}_i^{\mu\nu\rho\sigma}(p)
=\sum_{i=1}^3 {\cal T}_i^{\mu\nu\rho\sigma}(p)
g_{SSi}(p),
\eea
where $g_{SSi}=f_R^2 K^{(2)}_{Si}(y)$, 
with $K^{(2)}_{Si}$ defined in Appendix~\ref{ap_coilrodfies}. 
The tensors are
\bea
{\cal T}_1^{\mu\nu\rho\sigma}(p)&\equiv&
\delta^{\nu\rho}\delta^{\mu\sigma}+\delta^{\nu\sigma}\delta^{\mu\rho}
-\frac{14}{9}\delta^{\mu\nu}\delta^{\rho\sigma}\nonumber\\
&-&2\delta^{\nu\rho}\frac{p^\sigma p^\mu}{p^2}
-2\delta^{\nu\sigma}\frac{p^\rho p^\mu}{p^2}
-2\delta^{\mu\rho}\frac{p^\nu p^\sigma}{p^2}
-2\delta^{\mu\sigma}\frac{p^\rho p^\nu}{p^2}\nonumber\\
&+&\frac{8}{3}\delta^{\mu\nu}\frac{p^\rho p^\sigma}{p^2}
+\frac{8}{3}\delta^{\rho\sigma}\frac{p^\mu p^\nu}{p^2},\\
{\cal T}_2^{\mu\nu\rho\sigma}(p)&\equiv&
\delta^{\nu\rho}\frac{p^\sigma p^\mu}{p^2}
+\delta^{\nu\sigma}\frac{p^\rho p^\mu}{p^2}
+\delta^{\mu\rho}\frac{p^\sigma p^\nu}{p^2}
+\delta^{\mu\sigma}\frac{p^\rho p^\nu}{p^2}\nonumber\\
&-&\frac{4}{3}\delta^{\mu\nu}\frac{p ^\rho p^\sigma}{p^2}
-\frac{4}{3}\delta^{\sigma\rho}\frac{p^\mu p^\nu}{p^2}
+\frac{4}{9}\delta^{\mu\nu}\delta^{\rho\sigma},\\
{\cal T}_3^{\mu\nu\rho\sigma}(p)&\equiv&
\left(\frac{p^\mu p^\nu}{p^2}-\frac{\delta^{\mu\nu}}{3}\right)
\left(\frac{p^\rho p^\sigma}{p^2}-\frac{\delta^{\rho\sigma}}{3}\right),
\eea

Subsequently, the inverse of the matrix ${\bf W}$ is written as
\bea
{\bf \Gamma}^{\mu\nu\rho\sigma}\equiv \left(\begin{array}{ccc}
\Gamma^{(2)}_{RR}& \Gamma^{(2)}_{RC}& \Gamma^{(2)\rho\sigma}_{RS}\\
\Gamma^{(2)}_{CR}& \Gamma^{(2)}_{CC}& \Gamma^{(2)\rho\sigma}_{CS} \\
\Gamma^{(2)\mu\nu}_{SR} & \Gamma^{(2)\mu\nu}_{SC} & 
\Gamma^{(2)\mu\nu\rho\sigma}_{SS}
\end{array}\right),
\label{Gmatdef}
\eea
and the following notation is introduced:
\bea
W^{(2)}_{a b}=g_{a b},&\qquad& \Gamma^{(2)}_{a b}=h_{a b},\\
W^{(2)\mu\nu}_{c S}=\Delta^{\mu\nu} g_{c S},&\qquad&
\Gamma^{(2)\mu\nu}_{c S}=\Delta^{\mu\nu} h_{c S},\\
W^{(2)\mu\nu\rho\sigma}_{SS}=\sum_{i=1}^3 {\cal T}^{\mu\nu\rho\sigma}_i 
g_{SSi},&\qquad&
\Gamma^{(2)\mu\nu\rho\sigma}_{SS}=\sum_{i=1}^3 
{\cal T}^{\mu\nu\rho\sigma}_i h_{SSi}.
\eea
The $h$ functions are related to the $g$ functions by
\bea
h_{a b}&=&g^{-1}_{a b}-\frac{2}{3}g^{-1}_{a c}
g_{c S}h_{b S},\\
h_{a S}&=&\frac{3}{2}\frac{g_{bS}g^{-1}_{b a }}{(
g_{c S} g^{-1}_{c d} g_{d S}+5g_{SS1}-4g_{SS2}-g_{SS3})},\\
h_{SS1}&=&\frac{1}{4}g^{-1}_{SS1},\\
h_{SS2}&=&\frac{g^{-1}_{SS1} g_{SS2}}{4(g_{SS2}-g_{SS1})},\\
h_{SS3}&=&\frac{1}{2}\biggr(3g_{c S}h_{c S}+32g_{SS1}h_{SS1}
-16 g_{SS2}h_{SS1}\nonumber\\
&-&10 g_{SS3}h_{SS1}-16 g_{SS1} h_{SS2}+8 g_{SS2} h_{SS2}+8 g_{SS3} h_{SS2}
\biggr)\nonumber\\
&\times&
\frac{1}{5g_{SS1}-4g_{SS2}-g_{SS3}}.
\eea
Hence, we have obtained $\Gamma^{(2)}$.
\section{Vertices}\label{ap_vert}
In this appendix, all vertices up to fourth order which appear in the FHA
are given. First, the so called pure nematic vertices are discussed. 
\subsection{Nematic vertices}
The pure nematic vertices are those involving correlation functions of
the nematic order parameter $S^{\mu\nu}_R$.
Clearly, these  are $\gamma^{(2)}_{S S}$, 
$\gamma^{(3)}_{S S S}$, and $\gamma^{(4)}_{S S S S}$.
For instance, 
the vertex $\gamma^{(2)}_{SS}$ can be obtained from $\Gamma^{(2)\mu\nu\rho\sigma}_{SS}$
as defined in the previous appendix.

In what follows, the limit $p\rightarrow 0$ is considered, 
with $\vec p$ parallel to the orientational vector $\vec \eta$.\footnote{One 
can show that parallel alignment of $\vec p$ and $\vec \eta$ has 
highest instability temperature \cite{sigolifr94}.
In other words close to the spinodal, the angle $\theta$ 
between $\vec p$ and $\vec \eta$ is zero.}  
Consequently $\Delta^{\mu\nu}\rightarrow {\cal N}^{\mu\nu}$.
Thus
\bea
\gamma^{(2)}_{SS}(0)&=&{\cal N}^{\mu\nu}{\cal N}^{\rho\sigma}
\Gamma^{(2)\mu\nu\rho\sigma}_{SS}(0)\nonumber\\
&=& \sum_{i=1}^3 h_{SSi} {\cal T}_i^{\mu\nu\rho\sigma} 
{\cal N}^{\mu\nu}{\cal N}^{\rho\sigma}=
-\frac{20}{9}h_{SS1}(0)+\frac{16}{9}h_{SS2}(0)+\frac{4}{9}h_{SS3}(0)\nonumber\\
&=&-\lim_{p\rightarrow 0}\frac{1}{
(g_{C S}(p) g^{-1}_{C D}(p) g_{D S}(p)+5g_{SS1}(p)-4g_{SS2}(p)-g_{SS3}(p))}
=\frac{5}{f_R^2}.
\eea

The three-point nematic vertex is
\bea
\gamma^{(3)}_{SSS}(0,0)&=&{\cal N}{\cal N}{\cal N}\Gamma^{(3)}_{SSS}(0,0)
=-[{\cal N}\Gamma^{(2)}_{SS}(0)]^3 W^{(3)}_{SSS}(0,0)\nonumber\\
&=&-\left[\frac{15}{2 f_R^2} \right]^3 
{\cal N}{\cal N}{\cal N} W^{(3)}_{SSS}(0,0)\nonumber\\
&=&-\left[\frac{15}{2 f_R^2} \right]^3 f_R^3
\left(\frac{2}{3}\right)^3\frac{2}{35}=-\frac{1}{f_R^3}\frac{50}{7},
\eea
where we have used the identities
\bea
{\cal N}^{\mu\nu} \Gamma^{(2)\mu\nu\rho\sigma}_{SS}(0)=
\lim_{p\rightarrow 0}\sum_{i=1}^3 h_{SSi}(p) {\cal T}_i^{\mu\nu\rho\sigma}(p) 
{\cal N}^{\mu\nu}
=\frac{15}{2 f_R^2} {\cal N}^{\rho\sigma},
\eea
and
\bea
{\cal N}^{\mu\nu}{\cal N}^{\rho\sigma}
{\cal N}^{\kappa\lambda} W^{(3)\mu\nu\rho\sigma\kappa\lambda}_{SSS}(0,0)
&=&f_R^3
(2/3)^3\langle [P_2(\cos\theta)]^3\rangle_0\nonumber\\
&=&
f_R^3(2/3)^3\int\limits_{-1}^{1}dx\, [P_2(x)]^3/2
=f_R^3(2/3)^3(2/35).
\eea

Subsequently,
the four-point nematic vertex function can be computed,
\bea
\gamma^{(4)}_{SSSS}(0,0,0)&=&
{\cal N}{\cal N}{\cal N}{\cal N}\Gamma^{(4)}_{SSSS}(0,0,0)\nonumber\\
&=&-[{\cal N}\Gamma^{(2)}_{SS}(0)]^4 W^{(4)}_{SSSS}(0,0,0)\nonumber\\
&+&3  [{\cal N}\Gamma^{(2)}_{SS}(0)]^4 
W^{(3)}_{SSS}(0,0)\Gamma^{(2)}_{SS}(0) W^{(3)}_{SSS}(0,0)
\nonumber\\
&+& 3 [{\cal N}\Gamma^{(2)}_{SS}(0)]^4 W^{(2)}_{SS}(0) W^{(2)}_{SS}(0)
\nonumber\\
&=&\frac{1}{f_R^4} \frac{2550}{49},
\eea
where we have used that
\bea
{\cal N}{\cal N}{\cal N}{\cal N}W^{(4)}_{SSSS}(0,0,0)&=&
f_R^4 (2/3)^4 \langle [P_2(\cos\theta)]^4 \rangle =f_R^4 (2/3)^4 (3/35),\\
{\cal N}{\cal N}W^{(3)}_{SSS}(0,0)&=& f_R^3 (2/3)^2(2/35){\cal N},\label{wsss}\\
{\cal N}{\cal N} W^{(2)}_{SS}(0)&=&f_R^2 (2/3)^2(1/5).
\eea

\subsection{Density coefficients}
Now the expressions for the pure density vertices are listed.
The coefficient $\gamma^{(2)}_{\psi\psi}$ is given by
\bea
\gamma^{(2)}_{\psi\psi}(Q)=\epsilon_a \epsilon_b 
\Gamma^{(2)}_{a b}(Q)
=h_{RR}(Q)+h_{RR}(Q)-2h_{RC}(Q),\label{g2ppdef}
\eea
where the Roman indices $a,b$ sum over $R,C$.
At this point, we introduce the shorthand notation
\bea
\gamma_{a b}=\Gamma^{(2)}_{a b}(Q).\label{gamabdef}
\eea
The coefficient $\gamma^{(3)}_{\psi\psi\psi}$ is
\bea
\gamma^{(3)}_{\psi\psi\psi}(Q_1,Q_2)&=&\epsilon_a \epsilon_b \epsilon_c
\Gamma^{(3)}_{a b c}(Q_1,Q_2)\nonumber\\
&=&
-z_a(Q_1) z_b(Q_2) z_c(-Q_1-Q_2) W^{(3)}_{a b c} (Q_1,Q_2)
\eea
with $|Q_1|=|Q_2|=|Q_1+Q_2|$ (thus $\hat Q_1\cdot \hat Q_2=-1/2$) 
and where
\bea
z_a(Q)\equiv \epsilon_{a^\prime} \Gamma^{(2)}_{a^\prime a}(Q),
\eea
hence
\bea
z_R(Q)=h_{RR}(Q)-h_{RC}(Q),
\qquad 
z_C(Q)=h_{RC}(Q)-h_{CC}(Q).
\eea
The three point single chain correlation functions are
(again with $\hat Q_1\cdot \hat Q_2=-1/2$)
\bea
W^{(3)}_{RRR}(Q_1,Q_2)&=&w_{RRR},\quad w_{RRR}=f_R^3 K^{(3)}_{R1}(y,-1/2),\\
W^{(3)}_{RRC}(Q_1,Q_2)&=&w_{RRC},\quad w_{RRC}=f_R^2 f_C 
K^{(1)}_C(x) K^{(2)}_{R2}(y,-1/2),\\
W^{(3)}_{RCC}(Q_1,Q_2)&=&w_{RCC},\quad w_{RCC}=
f_R f_C^2 K^{(2)}_{C2}(x,-1/2) K^{(1)}_R(y),\\
W^{(3)}_{CCC}(Q_1,Q_2)&=&w_{CCC},\quad w_{CCC}=f_C^3 
K^{(3)}_{C1}(x,-1/2),
\eea
where $w_{RRC}=w_{RCR}=w_{CRR}$
and $w_{RCC}=w_{CCR}=w_{CRC}$.
Again the $K$ function are given in Appendix~\ref{ap_coilrodfies}.

The coefficient $\gamma^{(4)}_{\psi\psi\psi\psi}$
is (with $|Q_1|=|Q_2|=|Q_3|=|Q_1+Q_2+Q_3|$)
\bea
\gamma^{(4)}_{\psi\psi\psi\psi}(Q_1,Q_2,Q_3)&=&
\epsilon_a\epsilon_b\epsilon_c\epsilon_d
\Gamma^{(4)}_{abcd}(Q_1,Q_2,Q_3)\nonumber\\
&=&M_1(Q_1,Q_2,Q_3)
+M_2(Q_1,Q_2,Q_3)
\eea
where
\bea
M_1(Q_1,Q_2,Q_3)&=&-z_a(Q_1) z_b(Q_2) z_c(Q_3) z_d(-Q_1-Q_2-Q_3)
W^{(4)}_{abcd}(Q_1,Q_2,Q_3),\\
M_2(Q_1,Q_2,Q_3)&=&z_a(Q_1) z_b(Q_2) z_c(Q_3) z_d(-Q_1-Q_2-Q_3)\nonumber\\
&\times &\biggr[
W^{(3)}_{abe}(Q_1,Q_2)\Gamma^{(2)}_{ef}(-Q_1-Q_2)W^{(3)}_{fcd}(Q_1+Q_2,Q_3)
\nonumber\\
&+&W^{(3)}_{ace}(Q_1,Q_3)\Gamma^{(2)}_{ef}(-Q_1-Q_3)W^{(3)}_{fbd}(Q_1+Q_3,Q_2)
\nonumber\\
&+&
W^{(3)}_{ade}(Q_1,-Q_1-Q_2-Q_3)
\Gamma^{(2)}_{ef}(Q_2+Q_3)W^{(3)}_{fcb}(-Q_2-Q_3,Q_3)\nonumber\\
&+&\delta_K(Q_1+Q_2) W^{(2)}_{ab}(Q_1)W^{(2)}_{cd}(Q_3)
+\delta_K(Q_1+Q_3) W^{(2)}_{ac}(Q_1)W^{(2)}_{bd}(Q_2)\nonumber\\
&+&\delta_K(Q_2+Q_3) W^{(2)}_{ad}(Q_1)W^{(2)}_{bc}(Q_2)\biggr].
\label{M1M2expr}
\eea
Parts of the form
\bea
z_a z_b z_c z_d W^{(3)}_{abS}\Gamma^{(2)}_{SS}W^{(3)}_{cdS},
\eea
though formerly present in Eq.~(\ref{gam4ppppdef}),
are not included in Eq.~(\ref{M1M2expr}), 
since these parts can be shown to be negligible with respect to 
the other terms comprising the four point density vertex. 

We can compute the four-point single chain correlation functions.
Their contribution is decomposed as
\bea
W^{(4)}_{RRRR}(Q_1,Q_2,Q_3)&=&f_R^4
K^{(4)}_R(y,c_1,c_2),\quad y=\ell q^\ast,\\
W^{(4)}_{CRRR}(Q_1,Q_2,Q_3)&=&f_C f_R^3 K^{(1)}_{C}(x)K^{(3)}_{R2}(y,c_1,c_2),\\
W^{(4)}_{CCRR}(Q_1,Q_2,Q_3)&=&f_C^2 f_R^2 K^{(2)}_{C2}(x,c_1) 
K^{(2)}_{R2}(y,c_1), \\
W^{(4)}_{CCCR}(Q_1,Q_2,Q_3)&=&f_C^3 f_R K^{(3)}_{C2}(x,c_1,c_2)K^{(1)}_R(y),\\
W^{(4)}_{CCCC}(Q_1,Q_2,Q_3)&=&f_C^4 K^{(4)}_C(x,c_1,c_2),
\eea
where
\bea
c_1\equiv \frac{Q_1\cdot Q_2}{|Q|^2},\qquad
c_2\equiv \frac{Q_1\cdot Q_3}{|Q|^2},\qquad
c_3\equiv \frac{Q_2\cdot Q_3}{|Q|^2}=-1-c_1-c_2. \label{c1c2c3def}
\eea
Also, we have (with $c=Q_1\cdot Q_2/|Q|^2$)
\bea
W^{(3)}_{RRR}(Q_1,Q_2)&=&f_R^3 K^{(3)}_{R1}(y,c),\\
W^{(3)}_{RRC}(Q_1,Q_2)&=&f_R^2 f_C K^{(1)}_C(x(2+2c)) K^{(2)}_{R2}(y,c),\\
W^{(3)}_{RCR}(Q_1,Q_2)&=&f_R^2 f_C K^{(1)}_C(x) K^{(2)}_{R3}(y,c),\\
W^{(3)}_{CRR}(Q_1,Q_2)&=&f_R^2 f_C K^{(1)}_C(x) K^{(2)}_{R3}(y,c),\\
W^{(3)}_{RCC}(Q_1,Q_2)&=&f_R f_C^2 K^{(2)}_{C3}(x,c) K^{(1)}_R(y),\\
W^{(3)}_{CRC}(Q_1,Q_2)&=&f_R f_C^2 K^{(2)}_{C3}(x,c) K^{(1)}_R(y),\\
W^{(3)}_{CCR}(Q_1,Q_2)&=&f_R f_C^2 K^{(2)}_{C2}(x,c) K^{(1)}_R(y\sqrt{2+2c}),\\
W^{(3)}_{CCC}(Q_1,Q_2)&=&f_C^3 K^{(3)}_{C1}(x,c).
\eea

For the three morphologies it is sufficient to consider
four particular configurations of ``four vectors adding up to zero'' for 
$\gamma^{(4)}_{\psi\psi\psi\psi}(Q_1,Q_2,Q_3)$.
These configurations are, using Eqs.~(\ref{c1c2c3def}), 
\bea
c_1=-1,\quad c_2=-1,\quad c_3=1 \quad &\longrightarrow& \quad
\gamma^{(4)}_{\psi\psi\psi\psi1},\\
c_1=-1,\quad c_2=-1/2,\quad c_3=1/2 \quad &\longrightarrow& \quad
\gamma^{(4)}_{\psi\psi\psi\psi2},\\
c_1=-1,\quad c_2=0,\quad c_3=0 \quad &\longrightarrow& \quad
\gamma^{(4)}_{\psi\psi\psi\psi3},\\
c_1=-1/2,\quad c_2=-1/2,\quad c_3=0 \quad &\longrightarrow& \quad
\gamma^{(4)}_{\psi\psi\psi\psi4},
\eea
where $\gamma^{(4)}_{\psi\psi\psi\psi i}\equiv\gamma^{(4)}_{\psi\psi\psi\psi}(Q_1,Q_2,Q_3)$.

\subsection{Mixed vertices or coupling terms}
The most interesting vertices are those involving both nematic as well as density
order parameters, since these will give rise to coupling between nematic ordering
and microphase ordering.

The coupling vertex $\gamma^{(3)}_{\psi\psi S}$ is
\bea
\gamma^{(3)}_{\psi\psi S}(Q,-Q)&=&\epsilon_a \epsilon_b {\cal N} 
\Gamma^{(3)}_{abS}(Q,-Q)\nonumber\\
&=&-z_a(Q) z_b(Q) {\cal N}\Gamma^{(2)}_{SS}(0) W^{(3)}_{abS}(Q,-Q)\nonumber\\
&=&-z_a(Q) z_b(Q) {\cal N}\Gamma^{(2)}_{SS}(0) {\cal N}{\cal N}(3/2) 
W^{(3)}_{abS}(Q,-Q)\nonumber\\
&=&-\frac{2}{3} z_a(Q) z_b(Q) z_S  
\bar W^{(3)}_{abS}(Q,-Q),
\eea
where we have defined
\bea
z_S{\cal N}^{\mu\nu}{\cal N}^{\rho\lambda}\equiv \Gamma^{(2)\mu\nu\rho\lambda}_{SS}(0)
\quad \Longrightarrow \quad
z_S=\frac{45}{4f_R^2},
\eea
and
\bea
\bar W^{(3)}_{abS}(Q,-Q)={\cal N} W^{(3)}_{abS}(Q,-Q).
\eea

It is straightforward to show that
\bea
W^{(3)\mu\nu}_{RRS}(Q,-Q)&=&f_R W^{(2)\mu\nu}_{RS}(Q),\\
W^{(3)\mu\nu}_{RCS}(Q,-Q)&=&f_R W^{(2)\mu\nu}_{CS}(Q),\\
W^{(3)\mu\nu}_{CCS}(Q,-Q)&=&0.
\eea
Hence
\bea
\bar W^{(3)}_{RRS}(Q,-Q)&=&w_{RRS}P_2(\hat Q\cdot \eta),
\quad w_{RRS}=\frac{2}{3}f_R^3 K^{(2)}_{RS}(y),
\label{wrrs}\\
\bar W^{(3)}_{RCS}(Q,-Q)&=&w_{RCS}P_2(\hat Q\cdot \eta),
\quad w_{RCS}=\frac{2}{3}f_R^2f_C 
K^{(1)}_C(x) K^{(1)}_{S0}(y),
\label{wrcs}\\
W^{(3)\mu\nu}_{CCS}(Q,-Q)&=&0,\quad w_{CCS}=0,\label{wccs}
\eea
where $w_{RCS}=w_{CRS}$.

The four-point vertex $\gamma^{(4)}_{\psi\psi SS}$ reads 
\bea
\gamma^{(4)}_{\psi\psi SS}(Q,-Q,0)&=&\epsilon_a \epsilon_b {\cal N}{\cal N}
\Gamma^{(4)}_{a bSS}(Q,-Q,0)\nonumber\\
&=&-\left(\frac{2}{3}\right)^2
z_a(Q)z_b(Q)z_S^2\biggr[
\bar W^{(4)}_{ab SS}(Q,-Q,0)
-\bar W^{(2)}_{SS}(0) W^{(2)}_{ab}(Q)\nonumber\\
&-&z_S \bar W^{(3)}_{abS}(Q,-Q)\bar W^{(3)}_{SSS}(0,0)
-2\bar W^{(3)}_{aSc}(Q,0)\Gamma^{(2)}_{cd}(Q)\bar W^{(3)}_{dbS}(Q,-Q)
\biggr],\label{g4ppss}
\eea
with
\bea
\bar W^{(4)}_{ab SS}(Q,-Q,0)&=&{\cal N}{\cal N}W^{(4)}_{ab SS}(Q,-Q,0),\\
\bar W^{(2)}_{SS}(0)&=&{\cal N}{\cal N}W^{(2)}_{SS}(0),\\
\bar W^{(3)}_{abS}(Q,-Q)&=&{\cal N} W^{(3)}_{abS}(Q,-Q).
\eea
For instance, $W^{(4)}_{RR SS}$ is 
\bea
\bar W^{(4)}_{RR SS}(Q,-Q)=f_R^2 W^{(2)\mu\nu\rho\sigma}_{SS}(Q)
{\cal N}^{\mu\nu} {\cal N}^{\rho\sigma}
=f_R^4 \sum_{i=1}^3 K^{(2)}_{Si}(y) {\cal N}{\cal N} {\cal T}_i(Q),
\eea
and 
\bea
{\cal T}_1^{\mu\nu\rho\lambda}(Q){\cal N}^{\mu\nu}{\cal N}^{\rho\lambda}&=&-\frac{16}{9}
P_2\left(\hat Q\cdot \eta \right)-\frac{4}{9},\\
{\cal T}_2^{\mu\nu\rho\lambda}(Q){\cal N}^{\mu\nu}{\cal N}^{\rho\lambda}&=&\frac{8}{9}
P_2\left(\hat Q\cdot \eta\right)+\frac{8}{9},\\
{\cal T}_3^{\mu\nu\rho\lambda}(Q){\cal N}^{\mu\nu}{\cal N}^{\rho\lambda}&=&\frac{4}{9}
\left[P_2\left(\hat Q\cdot \eta\right)\right]^2.
\eea
Thus
\bea
\bar W^{(4)}_{RR SS}(Q,-Q)&=&w_{RRSS0}+w_{RRSS1}P_2(\hat Q\cdot \eta)
+w_{RRSS2}P_2^2(\hat Q\cdot \eta),\\
w_{RRSS0}&=&f_R^4\left(\frac{2}{3}\right)^2\left(-K^{(2)}_{S1}(y)+2K^{(2)}_{S2}(y)\right),\\
w_{RRSS1}&=&f_R^4\left(\frac{2}{3}\right)^2\left(-4K^{(2)}_{S1}(y)+2K^{(2)}_{S2}(y)\right),
\\
w_{RRSS2}&=&f_R^4\left(\frac{2}{3}\right)^2 K^{(2)}_{S3}(y).
\eea
Also
\bea
\bar W^{(4)}_{RC SS}(Q,-Q)&=&f_R^3 f_C K^{(1)}_C(x)
\sum_{i=1}^3 K^{(1)}_{Si}(y) {\cal N}{\cal N} {\cal T}_i(Q)\nonumber\\
&=&w_{RCSS0}+w_{RCSS1}P_2(\hat Q\cdot \eta)
+w_{RCSS2}P_2^2(\hat Q\cdot \eta),\\
w_{RCSS0}&=&f_Cf_R^3 K^{(1)}_C(x)
\left(\frac{2}{3}\right)^2\left(-K^{(1)}_{S1}(y)+2K^{(1)}_{S2}(y)\right),\\
w_{RCSS1}&=&f_Cf_R^3 K^{(1)}_C(x)
\left(\frac{2}{3}\right)^2\left(-4K^{(1)}_{S1}(y)+2K^{(1)}_{S2}(y)\right),
\\
w_{RCSS2}&=&f_Cf_R^3K^{(1)}_C(x)
\left(\frac{2}{3}\right)^2 K^{(1)}_{S3}(y).
\eea
and,
\bea
\bar W^{(4)}_{CC SS}(Q,-Q)&=&f_R^2 f_C^2 K^{(2)}_{C1}(x) \frac{4}{45}=w_{CCSS0}.
\eea
The three-point single chain correlation parts of Eq.~(\ref{g4ppss}) can 
be obtained 
from Eqs.~(\ref{wrrs})-(\ref{wccs}) and Eq.~(\ref{wsss}).

Subsequently the $\gamma^{(4)}_{\psi\psi S S}$ vertex is decomposed
in the following way,
\bea
\gamma^{(4)}_{\psi\psi S S}(Q,-Q,0)=\gamma^{(4)}_{\psi\psi S S 0}
+\gamma^{(4)}_{\psi\psi S S 1}P_2\left(\hat Q\cdot \eta \right)
+\gamma^{(4)}_{\psi\psi S S 2}P_2^2\left(\hat Q\cdot \eta \right),
\eea 
where
\bea
\gamma^{(4)}_{\psi\psi S S 0}&=&-\left(\frac{2}{3}\right)^2 z_S^2
\bigg[z_R^2w_{RRSS0}+2z_Rz_Cw_{RCSS0}+z_C^2w_{CCSS0}\nonumber\\
&&
-w_{SS}(z_R^2 w_{RR}+2z_R z_C w_{RC}+z_C^2 w_{CC})\bigg],\\
\gamma^{(4)}_{\psi\psi S S 1}&=&-\left(\frac{2}{3}\right)^2 z_S^2
\bigg[z_R^2w_{RRSS1}+2z_Rz_Cw_{RCSS1}\nonumber\\
&&
-z_S(z_R^2 w_{RRS}+2z_R z_C w_{RCS})w_{SSS}\bigg],\\
\gamma^{(4)}_{\psi\psi S S 2}&=&-\left(\frac{2}{3}\right)^2 z_S^2
\bigg[z_R^2w_{RRSS2}+2z_Rz_Cw_{RCSS2}\nonumber\\
&&
-2z_az_bw_{acS}\gamma_{cd}w_{dbS}\bigg].
\eea

Finally, we have the four-point vertex 
function $\gamma^{(4)}_{\psi\psi\psi S}$
\bea
\gamma^{(4)}_{\psi\psi\psi S}(Q_1,Q_2,Q_3)
&=&\epsilon_a \epsilon_b \epsilon_c {\cal N}\Gamma^{(4)}_{abc S}(Q_1,Q_2,Q_3)\nonumber\\
&=&-\frac{2}{3} z_a(Q_1) z_b(Q_2) z_c(Q_3) z_S\nonumber\\
&\times&
\biggr[\bar W^{(4)}_{abc S}(Q_1,Q_2,Q_3)\nonumber\\
&-&W^{(3)}_{abe}(Q_1,Q_2)\Gamma^{(2)}_{ef}(Q_3)\bar W^{(3)}_{fcS}(-Q_3,Q_3)
\nonumber\\
&-&W^{(3)}_{ace}(Q_1,Q_3)
\Gamma^{(2)}_{ef}(Q_2)\bar W^{(3)}_{fbS}(-Q_2,Q_2)
\nonumber\\
&-&\bar W^{(3)}_{aSe}(Q_1,0)
\Gamma^{(2)}_{ef}(-Q_1) W^{(3)}_{fcb}(Q_1,Q_3)\biggr],
\eea
with $Q_3=-Q_1-Q_2$.
In the FHA, the terms $W^{(2)}_{AB} W^{(2)}_{CS}$ vanish.

It can be shown that the correlation function $\bar W^{(4)}_{RRRS}$ is
\bea
\bar W^{(4)}_{RRRS}(Q_1,Q_2,Q_3)&=&
w_{RRRS}
\left[
P_2(\hat Q_1\cdot \eta)+
P_2(\hat Q_2\cdot \eta)+P_2(\hat Q_3\cdot \eta)
\right],\label{wrrrsdef}\\
w_{RRRS}&=&\frac{2}{3}f_R^4 K^{(3)}_S(y),
\eea
and
\bea
\bar W^{(4)}_{CRRS}(Q_1,Q_2,Q_3)
&+&\bar W^{(4)}_{RCRS}(Q_1,Q_2,Q_3)
+\bar W^{(4)}_{RRCS}(Q_1,Q_2,Q_3)
=w_{RRCS} \nonumber\\&&\times
\left[P_2(\hat Q_1\cdot \eta)+
P_2(\hat Q_2\cdot \eta)+P_2(\hat Q_3\cdot \eta)
\right],\label{wrrcsdef}\\
w_{RRCS}&=&\frac{2}{3}f_R^3 f_C K^{(1)}_C(x) K^{(2)}_{CS}(y).
\eea
The correlation function $\bar W^{(4)}_{CCRS}$ is (with $Q_3=-Q_1-Q_2$)
\bea
\bar W^{(4)}_{CCRS}(Q_1,Q_2,Q_3)
&+&\bar W^{(4)}_{CRCS}(Q_1,Q_2,Q_3)
+\bar W^{(4)}_{RCCS}(Q_1,Q_2,Q_3)
=w_{CCRS} \nonumber\\&&\times
\left[P_2(\hat Q_1\cdot \eta)+
P_2(\hat Q_2\cdot \eta)+P_2(\hat Q_3\cdot \eta)
\right],\label{wccrsdef}\\
w_{CCRS}&=&\frac{2}{3}f_R^2 f_C^2 K^{(2)}_{C2}(x) K^{(1)}_{S0}(y).
\eea
The correlation function $\bar W^{(4)}_{CCCS}$ vanishes in the FHA.

We can derive that, 
using Eq.~(\ref{wrrrsdef}), (\ref{wrrcsdef}) and (\ref{wccrsdef}),
the four-point vertex (which is fully symmetric in its arguments)
is
\bea
\gamma^{(4)}_{\psi\psi\psi S}(Q_1,Q_2,Q_3)
=\sigma_{\psi\psi\psi S}\left[
P_2(\hat Q_1\cdot \eta)+P_2(\hat Q_2\cdot \eta)+P_2(\hat Q_3\cdot \eta)
\right],
\eea
with $Q_3=-Q_1-Q_2$, and
where, 
\bea
\sigma_{\psi\psi\psi S}&=&-\frac{2}{3}z_S
\left[z_R^3 w_{RRRS}+z_R^2z_C w_{RRCS}+z_R z_C^2 w_{CCRS}\right]\nonumber\\
&+&\frac{2}{3}z_S \sum_{a,b,c,d,e}
z_a z_b z_c w_{abd} 
\gamma_{de}
w_{ec S},\label{sigpppsdef}
\eea
and $z_R=z_R(q^\ast)$, $z_C=z_C(q^\ast)$.
Note that indices $a,b,c,d,e=R,C$. 

\section{Definition of coil and rod functions}\label{ap_coilrodfies}
In this appendix, we list the so called coil and rod functions which
appear in single chain correlation functions and vertices as given in
the previous appendices.  The superscript of such a $K$ denotes the
number of integrations or internal points involved. For most $K$
functions an explicit expression can be given, however for certain rod
functions the integral form is retained, since no analytical form
could be obtained.

The coil functions are 
\bea
K^{(1)}_C(x)&=&\frac{1}{x}\left[1-{\rm e}^{-x}\right],\label{K1Cdef}\\
K^{(2)}_{C1}(x)
&=&f_D(x)=\frac{2}{x^2}\left[{\rm e}^{-x}+x-1\right],\label{debyedef}\\
K^{(2)}_{C2}(x)&=&2K^{(1)}_C(x)-K^{(2)}_{C1}(x),\label{gammaddef}
\eea
and
\bea
K^{(2)}_{C2}(x,c)&=&\int\limits_0^1d\tau_1\int\limits_0^1d\tau_2\,
{\rm e}^{-x(1+c)(2-\tau_1-\tau_2)+xc|\tau_1-\tau_2|}\nonumber\\
&=&\frac{{\rm e}^{-2(1+c)x}+(1+2c)
-2(1+c){\rm e}^{-x}}{(1+3c+2c^2)x^2},
\quad c\not=-\frac{1}{2},\,-1,
\label{K2C2def}\\
K^{(2)}_{C1}(x)&=&K^{(2)}_{C2}(x,-1)=f_D(x),\\
K^{(2)}_{C2}(x)&=&K^{(2)}_{C2}(x,-1/2)=
\frac{2}{x^2}\left[1-(1+x){\rm e}^{-x}\right],\\
K^{(2)}_{C3}(x,c)
&=&\int\limits_0^1d\tau_1\int\limits_0^1d\tau_2\,
{\rm e}^{-x(1+c)|\tau_1-\tau_2|-x(1+c)(1-\tau_2)+xc(1-\tau_1)}\nonumber\\
&=&\frac{{\rm e}^{-2(1+c)x}+(3+8c+4c^2)
-2(1+c)(2+x+2c(1+x)){\rm e}^{-x}}{2(1+c)(1+2c)x^2},\label{K2C3def}\\
&&\quad c\not=-\frac{1}{2},\,-1,\nonumber\\
K^{(2)}_{C3}(x,-1/2)&=&\frac{2}{x^2}\left[1-(1+x){\rm e}^{-x}\right],
\qquad K^{(2)}_{C3}(x,-1)=\frac{1}{x}\left[1-{\rm e}^{-x}\right].
\eea
The coil functions involving three or four integrals are 
\bea
K^{(3)}_{C1}(x,c)&=&\int\limits_0^1d\tau_1\int\limits_0^1d\tau_2
\int\limits_0^1d\tau_3\,
{\rm e}^{-x(1+c)|\tau_1-\tau_3|-x(1+c)|\tau_2-\tau_3|+xc|\tau_1-\tau_2|},
\label{K3C1def}\\
K^{(3)}_{C1}(x,-1/2)
&=&\frac{6}{x^3}\left[x(1+{\, \rm e}^{-x})-2(1-{\, \rm e}^{-x})\right].
\eea
and 
\bea
K^{(3)}_{C2}(x,c_1,c_2)&=&\int\limits_0^1d\tau_1\int\limits_0^1d\tau_2
\int\limits_0^1d\tau_3\,
{\rm e}^{x c_3(1-\tau_1+|\tau_2-\tau_3|)
+xc_2(1-\tau_2+|\tau_1-\tau_3|)
+xc_1(1-\tau_3+|\tau_1-\tau_2|)},\label{K3C2def}\\
K^{(4)}_C(x,c_1,c_2)&=&\int\limits_0^1d\tau_1\int\limits_0^1d\tau_2
\int\limits_0^1d\tau_3 \int\limits_0^1d\tau_4\,\nonumber\\
&\times&
{\rm e}^{xc_3(|\tau_1-\tau_4|+|\tau_2-\tau_3|)
+xc_2(|\tau_2-\tau_4|+|\tau_1-\tau_3|)
+xc_1(|\tau_3-\tau_4|+|\tau_1-\tau_2|)},\label{K4Cdef}
\eea
with $c_3=-1-c_1-c_2$.
Clearly explicit expressions for these integrals can be obtained,
these are already given elsewhere.

The particular rod functions of interest are
\bea
K^{(1)}_R(y)&=&\frac{{\rm Si}(y)}{y},\label{K1Rdef}\\
K^{(2)}_{R1}(y)&=&\frac{2}{y^2}\left[-1+\cos y
+y {\,\rm Si}(y)\right],\label{K2R1def}\\
K^{(2)}_{R2}(y,c)&=&\int\limits_0^1ds_1\int\limits_0^1ds_2\,
\frac{\sin y\sqrt{s_1^2+s_2^2+2s_1 s_2 c}}{y\sqrt{s_1^2+s_2^2+2s_1 s_2 c}},\label{K2R2def}
\\
K^{(2)}_{R3}(y,c)&=&\int\limits_0^1ds_1\int\limits_0^1ds_2\,
\frac{\sin y\sqrt{s_1^2+s_2^2(2+2c)-s_1 s_2 (2+2c)}}{
y\sqrt{s_1^2+s_2^2(2+2c)-s_1 s_2 (2+2c)}},\label{K2R3def}
\eea
and
\bea
K^{(3)}_{R1}(y,c)&=&\int\limits_0^1ds_1\int\limits_0^1ds_2\int\limits_0^1ds_3\,
\frac{\sin y\sqrt{\tau^{(3)}_{R1}}}{
y\sqrt{\tau^{(3)}_{R1}}},\label{K3R1}\\
\tau^{(3)}_{R1}&=&(s_1-s_3)^2+2(s_1-s_3)(s_2-s_3)c+(s_2-s_3)^2,\nonumber\\
K^{(3)}_{R2}(y,c_1,c_2)&\equiv& \int\limits_0^1ds_1
\int\limits_0^1ds_2\int\limits_0^1ds_3\,
\frac{\sin y\sqrt{\tau^{(3)}_{R2}}}{y\sqrt{\tau^{(3)}_{R2}}},\nonumber\\
\tau^{(3)}_{R2}&=&s_3^2+(s_1-s_3)^2+(s_2-s_3)^2-2s_3(s_1-s_3)c_1
\nonumber\\
&-&2s_3(s_2-s_3)c_2
-2(s_1-s_3)(s_2-s_3)(1+c_1+c_2),\label{K3R2def}\\
K^{(4)}_{R}(y,c_1,c_2)&\equiv& \int\limits_0^1ds_1
\int\limits_0^1ds_2\int\limits_0^1ds_3\int\limits_0^1ds_4\,
\frac{\sin y\sqrt{\tau^{(4)}_{R}}}{y\sqrt{\tau^{(4)}_{R}}},\nonumber\\
\tau^{(4)}_{R}&=&(s_1-s_4)^2+(s_2-s_4)^2+(s_3-s_4)^2+2(s_1-s_4)(s_2-s_4)c_1
\nonumber\\
&+&2(s_1-s_4)(s_3-s_4)c_2
-2(s_2-s_4)(s_3-s_4)(1+c_1+c_2).\label{K4Rdef}
\eea

Also, there are the so called mixed terms
\bea
K^{(2)}_{RS}(y)&=&\frac{1}{y^3}
\left[4y-y\cos y-3\sin y-y^2{\,\rm Si}(y)
\right],\label{K2RSdef}\\
K^{(2)}_{CS}(y)&=&\int\limits_0^1ds_1\int\limits_0^1ds_2\,
\frac{s_2(2s_2-s_1)}{\tau^{(2)5}_{CS} y^3}\left[
(y^2\tau^{(2)2}_{CS}-3)\sin (y\tau^{(2)}_{CS})+3 y\tau^{(2)}_{CS} 
\cos(y\tau^{(2)}_{CS})\right],\nonumber\\
\tau^{(2)}_{CS}&=&\sqrt{s_1^2+s_2^2-s_1 s_2}.\label{K2CSdef}
\eea

Finally, we list the $K$ function which appear as factors
in vertices with nematic order parameters,
\bea
K^{(1)}_{S0}(y)&\equiv&
\int_0^1\frac{ds}{s^2}\,
\left(-\frac{d^2}{dy^2}+\frac{1}{y}\frac{d}{dy}\right)\frac{\sin ys}{ys}\nonumber\\
&=&\left(-\frac{1}{2y^3}\right)
\left[3y\cos y-3\sin y+y^2{\,\rm Si}(y)\right],\label{K1S0def}\\
K^{(1)}_{S1}(y)
&\equiv&
\int_0^1\frac{ds}{s^4}\,\left(\frac{1}{y^2}\frac{d^2}{dy^2}
-\frac{1}{y^3}\frac{d}{dy}\right)\frac{\sin ys}{ys}\nonumber\\
&=&\frac{1}{8y^5}\left[y^4 {\rm Si}(y)+y(y^2+6)\cos y+(y^2-6)\sin y\right],\label{K1S1def}\\
K^{(1)}_{S2}(y)&\equiv&
\int_0^1\frac{ds}{s^4}\,
\left(\frac{1}{y}\frac{d^3}{dy^3}-\frac{1}{y^2}\frac{d^2}{dy^2}
+\frac{1}{y^3}\frac{d}{dy}
\right)
\frac{\sin ys}{ys}\nonumber\\
&=&\frac{1}{8y^5}\left[y^4 {\rm Si}(y)+y(y^2-18)\cos y
+(18-7 y^2)\sin y\right],\label{K1S2def}\\
K^{(1)}_{S3}(y)&\equiv&
\int_0^1\frac{ds}{s^4}\,
\left(\frac{d^4}{dy^4}-\frac{6}{y}\frac{d^3}{dy^3}
+\frac{15}{y^2}\frac{d^2}{dy^2}-\frac{15}{y}\frac{d}{dy}
\right)
\frac{\sin ys}{ys}\nonumber\\
&=&
\frac{1}{8y^5}
\left[3y^3 {\rm Si}(y)-5y(y^2-42)\cos y+15(5y^2-14)\sin y\right],
\label{K1S3def}
\eea
and
\bea
K^{(2)}_{S1}(y)&\equiv&\int_0^1\int_0^1\frac{dsds'}{(s-s')^4}\,
\left(\frac{1}{y^2}\frac{d^2}{dy^2}-\frac{1}{y^3}\frac{d}{dy}\right)\frac{\sin y(s-s')}{y(s-s')}\nonumber\\
&=&\left(\frac{1}{12 y^5}\right)
\biggr[3y^4{\,\rm Si}(y)-8y^3+3y(y^2-2)\cos y +3(y^2+2) \sin y 
\biggr],\label{K2S1def}\\
K^{(2)}_{S2}(y)&\equiv&\int_0^1\int_0^1\frac{dsds'}{(s-s')^4}\,
\left(\frac{1}{y}\frac{d^3}{dy^3}-\frac{1}{y^2}\frac{d^2}{dy^2}
+\frac{1}{y^3}\frac{d}{dy}
\right)\frac{\sin y(s-s')}{y(s-s')}\nonumber\\
&=&
\left(
\frac{1}{4y^5}\right)\biggr[y^4{\,\rm Si}(y)
+y(y^2+6)\cos y + (y^2-6)\sin y 
\biggr],\label{K2S2def}\\
K^{(2)}_{S3}(y)&\equiv&
\int_0^1\int_0^1\frac{dsds'}{(s-s')^4}\,
\left(\frac{d^4}{dy^4}-\frac{6}{y}\frac{d^3}{dy^3}
+\frac{15}{y^2}\frac{d^2}{dy^2}-\frac{15}{y}\frac{d}{dy}
\right)\frac{\sin y(s-s')}{y(s-s')}\nonumber\\
&=&\left(\frac{1}{12y^5}\right)
\biggr[9y^4{\,\rm Si}(y)-64y^3
+y(9y^2-210)\cos y -15(y^2-14)\sin y 
\biggr],\label{K2S3def}
\eea
and
\bea
K^{(3)}_S(y)&=&\int\limits_0^1ds_1\int\limits_0^1ds_2\int\limits_0^1 ds_3\,
\frac{(s_1-s_3)(s_2-s_3)}{\tau^{(3)5}_S y^3}\left[
(y^2\tau^{(3)2}_S-3)\sin (y\tau^{(3)}_S)+3 y\tau^{(3)}_S 
\cos(y\tau^{(3)}_S)\right],\nonumber\\
\tau^{(3)}_S&=&\sqrt{s_1^2+s_2^2+s_3^2-s_1s_2-s_1s_3-s_2s_3}.\label{K3Sdef}
\eea
\newif\ifdraft
\newif\ifaps
\def\optbibfield#1{\ifdraft{#1}\else\ifaps\else{#1}\fi\fi}
%

%
\end{document}